\documentclass{article}
\usepackage{graphicx}
\usepackage{amsmath}

\newtheorem{theorem}{Theorem}

\newtheorem{remark}[theorem]{Remark}

\input{tcilatex}
\begin{document}

\title{ A Mechanical Implementation and Diagrammatic Calculation of
Entangled Basis States}
\author{F.A. Buot,$^{1,2,3}$A.R. Elnar,$^{2}$ G. Maglasang,$^{2}$ and C.M.
Galon$^{2}$ \\
%EndAName
$^{1}$C\&LB Research Institute, Carmen, Cebu 6005, Philippines\\
$^{2}$CTCMP, Cebu Normal University, Cebu City 6000, Philippines\\
$^{3}$LCFMNN, Department of Physics, University of San Carlos, \\
Talamban, Cebu 6000, Philippines}
\maketitle

\begin{abstract}
We give for the first time a diagrammatic calculational tool of quantum
entanglement. We present a pedagogical and simple mechanical implementation
of quantum entanglement or "spooky action at a distance" to give a tangible
realization of this weird quantum mechanical concept alien to classical
physics. When two or more particles are correlated in a certain way, no
matter how far apart they are in space, their states remain correlated.
Their correlation, which is instantaneous, does not seem to involve any
communication which is limited by the speed of light. The same mechanical
implementation demonstrates the fundamental physical limits of any
computational processes. The analytical derivations of calculational
entangled basis states are given and their corresponding diagrammatic
representations give an efficient aid in determining the calculational
entangled basis states. A quantum Fourier transform for the two-state
diagrams representing entangled basis states ('renormalized qubits') can
also be formulated. Our results seem to advocate the idea that quantum
entanglement generates the extra dimensions of the gravitational theory,
indeed quantum entanglement is related to deep issues in the unification of
general relativity and quantum mechanics. This extra dimensions of spacetime
entanglement are currently being speculated in the literature.
\end{abstract}

\section{INTRODUCTION}

The principal characteristics of quantum mechanics, which are alien to
classical physics, are as follows. First is the notion of \textit{uncertainty%
}, i.e., not all the classical physical observable properties of a system
can be simultaneously determined with exact precision. For example, position
and momentum cannot both be known at the same time. There may other sets of
observable properties that share the uncertainty properties. Second, the
notion of \textit{superposition}, for example, a cat is a superposition of
alive and dead cat. In other words an arbitrary quantum state is a
superposition of several quantum sates that describe measurable events.
Third, the notion of \textit{entanglement}, for example, if two photons or
two particle spins become entangled, i.e., they are allowed to interact
initially so that they will subsequently be defined by a single quantum
state, then once they are separated, they will still share the same quantum
state. So measuring one will determine the state of the other: for example,
with a spin-zero entangled state, if one particle is measured to be in a
spin-up state, the other is instantly forced to be in a spin-down state. In
this paper, we would like to address the weirdest aspect of quantum
mechanics, i.e., the notion of quantum entanglement.

Today, quantum entanglement forms the basis of several quantum technologies.
In quantum teleportation and quantum cryptography, entangled particles are
used to transmit signals that cannot be intercepted by an eavesdropper. The
preliminary viable quantum cryptography systems are now being used by banks.
And the exploding field of quantum computation uses superposition and
entangled quantum states to perform computational calculations in parallel
and at ultra-speed, so that some types of classically impractical
calculations can be done by quantum computers in reasonable length of time.

Historically, in a 1935 paper, Einstein, Boris Podolsky and Nathan Rosen
argue that quantum mechanics was not a complete physical theory. Known today
as the \textquotedblleft EPR paradox,\textquotedblright\ the thought
experiment was meant to demonstrate the inherent conceptual, in terms of
classical analogies, of quantum theory. Entanglement claims that the result
of a measurement on one particle of an entangled quantum system can have an
instantaneous effect on another particle, regardless of their distance. But
the EPR paradox did help deepen our understanding of quantum mechanics by
exposing the fundamentally non-classical characteristics of the measurement
process. Before that paper, most physicists viewed a measurement as a
physical disturbance inflicted directly on the measured system: one shines
light onto an electron to determine its position, but this disturbs the
electron and produces uncertainties. The EPR paradox shows that a
\textquotedblleft measurement\textquotedblright\ can be performed on a
particle without disturbing it directly, by performing a measurement on a
distant entangled particle.

Einstein postulated the existence of hidden variables, yet unknown local
properties of the system which should account for the discrepancy, so that
no instantaneous spooky action would be necessary. Bohr disagreed vehemently
with this view and defended the far stricter Copenhagen interpretation of
quantum mechanics. The two men were passionate about the subject, especially
at the Solvay Conferences of 1927 and 1930, neither Einstein nor Bohr
conceded defeat.

In this paper, we try to propose a simple mechanical model of a classical
analogy to the 'spooky at a distance' property of quantum entanglement. This
is a simple mechanical model, an ideal classical analogy for Einstein
arguments. For what it is worth, it gives a simple classical/mechanical
implementation and pedagogical value to the concept of quantum entanglement,
which may revive interest in some intermediary medium (not hidden variables,
but part of the whole system, i.e., nonlocality and hence does not violate
Bell's theorem) in quantum entanglement.

\section{ANALYTICAL DERIVATION OF ENTANGLED BASIS STATES}

First, we give \ formal derivation of calculational entangled basis states
as our foundation to the diagrammatic techniques. This was treated before in
the author's book \cite{buotbook}. Since the treatment there is not widely
annunciated, we give it here for completeness. Consider the identity 
\begin{equation*}
\left\vert p\right\rangle =\sum\limits_{q}\left\langle q\right\vert \left.
p\right\rangle \ \left\vert q\right\rangle ,
\end{equation*}%
where the $\left\langle q\right\vert \left. p\right\rangle $ is the
transformation function. For discrete quantum mechanics, this is given by
the discrete Fourier transform function, 
\begin{equation*}
\left\langle q\right\vert \left. p\right\rangle =\frac{1}{\sqrt{N}}\exp
\left( -\frac{i}{\hbar }p\cdot q\right) .
\end{equation*}%
Therefore,%
\begin{equation*}
\left\vert p\right\rangle =\frac{1}{\sqrt{N}}\sum\limits_{q}\exp \left( -%
\frac{i}{\hbar }p\cdot q\right) \ \left\vert q\right\rangle .
\end{equation*}%
The product state in the 'momentum' basis is expanded in terms of the
'position' basis%
\begin{eqnarray*}
\left\vert p^{\prime }\right\rangle \left\vert p^{\prime \prime
}\right\rangle &=&\frac{1}{\sqrt{N}}\sum\limits_{q^{\prime }}\exp \left( -%
\frac{i}{\hbar }p^{\prime }\cdot q^{\prime }\right) \ \left\vert q^{\prime
}\right\rangle \frac{1}{\sqrt{N}}\sum\limits_{q^{\prime \prime }}\exp \left(
-\frac{i}{\hbar }p^{\prime \prime }\cdot q^{\prime \prime }\right) \
\left\vert q^{\prime \prime }\right\rangle \\
&=&\frac{1}{N}\sum\limits_{q^{\prime },q^{\prime \prime }}\exp \left( -\frac{%
i}{\hbar }p^{\prime }\cdot q^{\prime }\right) \exp \left( -\frac{i}{\hbar }%
p^{\prime \prime }\cdot q^{\prime \prime }\right) \ \left\vert q^{\prime
}\right\rangle \ \left\vert q^{\prime \prime }\right\rangle .
\end{eqnarray*}%
We can expressed the right hand side in terms of 'correlated' (entangled)
product states by writing, $q^{\prime \prime }=q^{\prime }+m$, where $m$ is
the quantum-correlation 'distance', 
\begin{eqnarray*}
\left\vert p^{\prime }\right\rangle \left\vert p^{\prime \prime
}\right\rangle &=&\frac{1}{N}\sum\limits_{q^{\prime },m}\exp \left( -\frac{i%
}{\hbar }p^{\prime }\cdot q^{\prime }\right) \exp \left( -\frac{i}{\hbar }%
p^{\prime \prime }\cdot \left( q^{\prime }+m\right) \right) \ \left\vert
q^{\prime }\right\rangle \ \left\vert q^{\prime }+m\right\rangle \\
&=&\frac{1}{N}\sum\limits_{q^{\prime },m}\exp \left( -\frac{i}{\hbar }\left(
p^{\prime }+p^{\prime \prime }\right) \cdot q^{\prime }\right) \exp \left( -%
\frac{i}{\hbar }p^{\prime \prime }\cdot m\right) \ \left\vert q^{\prime
}\right\rangle \ \left\vert q^{\prime }+m\right\rangle .
\end{eqnarray*}%
Now writing $p^{\prime }+p^{\prime \prime }=p$, we have%
\begin{eqnarray*}
\left\vert p^{\prime }\right\rangle \left\vert p-p^{\prime }\right\rangle &=&%
\frac{1}{N}\sum\limits_{q^{\prime },m}\exp \left( -\frac{i}{\hbar }\left(
p\right) \cdot q^{\prime }\right) \exp \left( -\frac{i}{\hbar }\left(
p-p^{\prime }\right) \cdot m\right) \ \left\vert q^{\prime }\right\rangle \
\left\vert q^{\prime }+m\right\rangle \\
&=&\frac{1}{\sqrt{N}}\sum\limits_{m}\exp \left( -\frac{i}{\hbar }\left(
p-p^{\prime }\right) \cdot m\right) \left\{ \frac{1}{\sqrt{N}}%
\sum\limits_{q^{\prime }}\exp \left( -\frac{i}{\hbar }p\cdot q^{\prime
}\right) \ \left\vert q^{\prime }\right\rangle \ \left\vert q^{\prime
}+m\right\rangle \right\} \\
&=&\frac{1}{\sqrt{N}}\sum\limits_{m}\exp \left( -\frac{i}{\hbar }\left(
p-p^{\prime }\right) \cdot m\right) \left\vert \psi _{p,m}\right\rangle .
\end{eqnarray*}%
The inverse transformation gives $\left\vert \psi _{p,m}\right\rangle $ in
terms of the the 'momentum' basis products%
\begin{equation}
\left\vert \psi _{p,m}\right\rangle =\frac{1}{\sqrt{N}}\sum\limits_{p^{%
\prime }}\exp \left( \frac{i}{\hbar }\left( p-p^{\prime }\right) \cdot
m\right) \left\vert p^{\prime }\right\rangle \left\vert p-p^{\prime
}\right\rangle .  \label{inverse_bell-momentum_basis}
\end{equation}%
Clearly, the correlated basis defined by $\left\vert \psi
_{p,m}\right\rangle $ forms orthonormal and complete set.

\subsection{Bell Basis}

Thus, for $N=2$, we have the standard momentum product states in terms of
the so-called Bell basis, $\left\vert \psi _{p,m}\right\rangle $, for
example, 
\begin{equation*}
\left\vert k^{\prime }\right\rangle \left\vert k-k^{\prime }\right\rangle =%
\frac{1}{\sqrt{2}}\sum\limits_{m=0}^{1}\exp \left( -\pi i\left( k-k^{\prime
}\right) \cdot m\right) \left\vert \psi _{k,m}\right\rangle ,
\end{equation*}%
which yields%
\begin{equation*}
\left\vert 0\right\rangle \left\vert k\right\rangle =\frac{1}{\sqrt{2}}%
\sum\limits_{m=0}^{1}\exp \left( -\pi ik\cdot m\right) \left\vert \psi
_{k,m}\right\rangle ,
\end{equation*}%
and using the identities%
\begin{eqnarray*}
\left\vert \psi _{00}\right\rangle &=&\left\vert \Phi ^{+}\right\rangle , \\
\left\vert \psi _{01}\right\rangle &=&\left\vert \Psi ^{+}\right\rangle , \\
\left\vert \psi _{10}\right\rangle &=&\left\vert \Phi ^{-}\right\rangle , \\
\left\vert \psi _{11}\right\rangle &=&\left\vert \Psi ^{-}\right\rangle ,
\end{eqnarray*}%
we have,%
\begin{eqnarray*}
\left\vert 0\right\rangle \left\vert 0\right\rangle &=&\frac{1}{\sqrt{2}}%
\sum\limits_{m=0}^{1}\exp \left( -\pi ik\cdot m\right) \left\vert \psi
_{0,m}\right\rangle \\
&=&\frac{1}{\sqrt{2}}\left( \left\vert \psi _{00}\right\rangle +\left\vert
\psi _{01}\right\rangle \right) \\
&=&\frac{1}{\sqrt{2}}\left( \left\vert \Phi ^{+}\right\rangle +\left\vert
\Psi ^{+}\right\rangle \right) ,
\end{eqnarray*}%
\begin{eqnarray*}
\left\vert 0\right\rangle \left\vert 1\right\rangle &=&\frac{1}{\sqrt{2}}%
\sum\limits_{m=0}^{1}\exp \left( -\pi i1\cdot m\right) \left\vert \psi
_{1,m}\right\rangle \\
&=&\frac{1}{\sqrt{2}}\left( \left\vert \psi _{1,0}\right\rangle -\left\vert
\psi _{1,1}\right\rangle \right) \\
&=&\frac{1}{\sqrt{2}}\left( \left\vert \Phi ^{-}\right\rangle -\left\vert
\Psi ^{-}\right\rangle \right) ,
\end{eqnarray*}%
We also have%
\begin{equation*}
\left\vert 1\right\rangle \left\vert k+1\right\rangle =\frac{1}{\sqrt{2}}%
\sum\limits_{m=0}^{1}\exp \left( -\pi i\left( k+1\right) \cdot m\right)
\left\vert \psi _{k,m}\right\rangle \left( \func{mod}2\right) ,
\end{equation*}%
which yields%
\begin{eqnarray*}
\left\vert 1\right\rangle \left\vert 1\right\rangle &=&\frac{1}{\sqrt{2}}%
\sum\limits_{m=0}^{1}\exp \left( -\pi i\left( 1\right) \cdot m\right)
\left\vert \psi _{0,m}\right\rangle \\
&=&\frac{1}{\sqrt{2}}\left( \left\vert \psi _{0,0}\right\rangle -\left\vert
\psi _{0,1}\right\rangle \right) \\
&=&\frac{1}{\sqrt{2}}\left( \left\vert \Phi ^{+}\right\rangle -\left\vert
\Psi ^{+}\right\rangle \right) ,
\end{eqnarray*}%
\begin{eqnarray*}
\left\vert 1\right\rangle \left\vert 1+1\right\rangle &=&\frac{1}{\sqrt{2}}%
\sum\limits_{m=0}^{1}\exp \left( -\pi i\left( 1+1\right) \cdot m\right)
\left\vert \psi _{1,m}\right\rangle \left( \func{mod}2\right) , \\
\left\vert 1\right\rangle \left\vert 0\right\rangle &=&\frac{1}{\sqrt{2}}%
\left( \left\vert \psi _{1,0}\right\rangle +\left\vert \psi
_{1,1}\right\rangle \right) \\
&=&\frac{1}{\sqrt{2}}\left( \left\vert \Phi ^{-}\right\rangle +\left\vert
\Psi ^{-}\right\rangle \right) .
\end{eqnarray*}%
Therefore, for $N=2$, we have the following transformation from the
maximally entangled Bell basis to the standard 'momentum'-state product
basis given by%
\begin{equation*}
\left( 
\begin{array}{c}
\left\vert 0\right\rangle \left\vert 0\right\rangle \\ 
\left\vert 0\right\rangle \left\vert 1\right\rangle \\ 
\left\vert 1\right\rangle \left\vert 0\right\rangle \\ 
\left\vert 1\right\rangle \left\vert 1\right\rangle%
\end{array}%
\right) _{p}=\frac{1}{\sqrt{2}}\left( 
\begin{array}{cccc}
1 & 1 & 0 & 0 \\ 
0 & 0 & 1 & -1 \\ 
0 & 0 & 1 & 1 \\ 
1 & -1 & 0 & 0%
\end{array}%
\right) \left( 
\begin{array}{c}
\left\vert \Phi ^{+}\right\rangle \\ 
\left\vert \Psi ^{+}\right\rangle \\ 
\left\vert \Phi ^{-}\right\rangle \\ 
\left\vert \Psi ^{-}\right\rangle%
\end{array}%
\right) _{Bell}.
\end{equation*}%
Hence,%
\begin{eqnarray}
\left( 
\begin{array}{c}
\left\vert \Phi ^{+}\right\rangle \\ 
\left\vert \Psi ^{+}\right\rangle \\ 
\left\vert \Phi ^{-}\right\rangle \\ 
\left\vert \Psi ^{-}\right\rangle%
\end{array}%
\right) _{Bell} &=&\left[ \frac{1}{\sqrt{2}}\left( 
\begin{array}{cccc}
1 & 1 & 0 & 0 \\ 
0 & 0 & 1 & -1 \\ 
0 & 0 & 1 & 1 \\ 
1 & -1 & 0 & 0%
\end{array}%
\right) \right] ^{-1}\left( 
\begin{array}{c}
\left\vert 0\right\rangle \left\vert 0\right\rangle \\ 
\left\vert 0\right\rangle \left\vert 1\right\rangle \\ 
\left\vert 1\right\rangle \left\vert 0\right\rangle \\ 
\left\vert 1\right\rangle \left\vert 1\right\rangle%
\end{array}%
\right) _{p}  \notag \\
&=&\frac{1}{\sqrt{2}}\left( 
\begin{array}{cccc}
1 & 0 & 0 & 1 \\ 
1 & 0 & 0 & -1 \\ 
0 & 1 & 1 & 0 \\ 
0 & -1 & 1 & 0%
\end{array}%
\right) \left( 
\begin{array}{c}
\left\vert 0\right\rangle \left\vert 0\right\rangle \\ 
\left\vert 0\right\rangle \left\vert 1\right\rangle \\ 
\left\vert 1\right\rangle \left\vert 0\right\rangle \\ 
\left\vert 1\right\rangle \left\vert 1\right\rangle%
\end{array}%
\right) _{p}.  \label{inverse-mat_bell-mom_basis}
\end{eqnarray}%
The above result also follows from the general inverse expression, Eq. (\ref%
{inverse_bell-momentum_basis}), in terms of 'momentum' product basis,%
\begin{equation*}
\left\vert \psi _{p,m}\right\rangle =\frac{1}{\sqrt{N}}\sum\limits_{p^{%
\prime }}\exp \left( \frac{i}{\hbar }\left( p-p^{\prime }\right) \cdot
m\right) \left\vert p^{\prime }\right\rangle \left\vert p-p^{\prime
}\right\rangle ,
\end{equation*}%
so that for $N=2$,%
\begin{equation*}
\left\vert \psi _{k,m}\right\rangle =\frac{1}{\sqrt{2}}\sum\limits_{k^{%
\prime }}\exp \left( i\pi \left( k-k^{\prime }\right) \cdot m\right)
\left\vert k^{\prime }\right\rangle \left\vert k-k^{\prime }\right\rangle ,
\end{equation*}%
which yields 
\begin{eqnarray*}
\left\vert \psi _{00}\right\rangle &=&\left\vert \Phi ^{+}\right\rangle =%
\frac{1}{\sqrt{2}}\sum\limits_{k^{\prime }}\exp \left( i\pi \left(
0-k^{\prime }\right) \cdot 0\right) \left\vert k^{\prime }\right\rangle
\left\vert 0-k^{\prime }\right\rangle \\
&=&\frac{1}{\sqrt{2}}\left( \left\vert 0\right\rangle \left\vert
0\right\rangle +\left\vert 1\right\rangle \left\vert 1\right\rangle \right) ,
\end{eqnarray*}%
\begin{eqnarray*}
\left\vert \psi _{01}\right\rangle &=&\left\vert \Psi ^{+}\right\rangle =%
\frac{1}{\sqrt{2}}\sum\limits_{k^{\prime }}\exp \left( i\pi \left( k^{\prime
}\right) \cdot 1\right) \left\vert k^{\prime }\right\rangle \left\vert
k^{\prime }\right\rangle \\
&=&\frac{1}{\sqrt{2}}\left( \left\vert 0\right\rangle \left\vert
0\right\rangle -\left\vert 1\right\rangle \left\vert 1\right\rangle \right) ,
\end{eqnarray*}%
\begin{eqnarray*}
\left\vert \psi _{10}\right\rangle &=&\left\vert \Phi ^{-}\right\rangle =%
\frac{1}{\sqrt{2}}\sum\limits_{k^{\prime }}\exp \left( i\pi \left(
1-k^{\prime }\right) \cdot 0\right) \left\vert k^{\prime }\right\rangle
\left\vert 1-k^{\prime }\right\rangle \\
&=&\frac{1}{\sqrt{2}}\left( \left\vert 0\right\rangle \left\vert
1\right\rangle +\left\vert 1\right\rangle \left\vert 0\right\rangle \right) ,
\end{eqnarray*}%
\begin{eqnarray*}
\left\vert \psi _{1,1}\right\rangle &=&\left\vert \psi _{k,m}\right\rangle =%
\frac{1}{\sqrt{2}}\sum\limits_{k^{\prime }}\exp \left( i\pi \left(
1-k^{\prime }\right) \cdot 1\right) \left\vert k^{\prime }\right\rangle
\left\vert 1-k^{\prime }\right\rangle \\
&=&-\frac{1}{\sqrt{2}}\left( \left\vert 0\right\rangle \left\vert
1\right\rangle -\left\vert 1\right\rangle \left\vert 0\right\rangle \right) .
\end{eqnarray*}%
We also have%
\begin{equation}
\left( 
\begin{array}{c}
\left\vert \Phi ^{+}\right\rangle \\ 
\left\vert \Psi ^{+}\right\rangle \\ 
\left\vert \Phi ^{-}\right\rangle \\ 
\left\vert \Psi ^{-}\right\rangle%
\end{array}%
\right) _{Bell}=\frac{1}{\sqrt{2}}\left( 
\begin{array}{cccc}
1 & 0 & 0 & 1 \\ 
0 & 1 & 1 & 0 \\ 
1 & 0 & 0 & -1 \\ 
0 & 1 & -1 & 0%
\end{array}%
\right) \left( 
\begin{array}{c}
\left\vert 0\right\rangle \left\vert 0\right\rangle \\ 
\left\vert 0\right\rangle \left\vert 1\right\rangle \\ 
\left\vert 1\right\rangle \left\vert 0\right\rangle \\ 
\left\vert 1\right\rangle \left\vert 1\right\rangle%
\end{array}%
\right) _{q}.  \label{inverse-mat_bell-pos_basis}
\end{equation}%
Therefore, we have the identity%
\begin{equation*}
\frac{1}{\sqrt{2}}\left( 
\begin{array}{cccc}
1 & 0 & 0 & 1 \\ 
1 & 0 & 0 & -1 \\ 
0 & 1 & 1 & 0 \\ 
0 & -1 & 1 & 0%
\end{array}%
\right) \left( 
\begin{array}{c}
\left\vert 0\right\rangle \left\vert 0\right\rangle \\ 
\left\vert 0\right\rangle \left\vert 1\right\rangle \\ 
\left\vert 1\right\rangle \left\vert 0\right\rangle \\ 
\left\vert 1\right\rangle \left\vert 1\right\rangle%
\end{array}%
\right) _{p}=\frac{1}{\sqrt{2}}\left( 
\begin{array}{cccc}
1 & 0 & 0 & 1 \\ 
0 & 1 & 1 & 0 \\ 
1 & 0 & 0 & -1 \\ 
0 & 1 & -1 & 0%
\end{array}%
\right) \left( 
\begin{array}{c}
\left\vert 0\right\rangle \left\vert 0\right\rangle \\ 
\left\vert 0\right\rangle \left\vert 1\right\rangle \\ 
\left\vert 1\right\rangle \left\vert 0\right\rangle \\ 
\left\vert 1\right\rangle \left\vert 1\right\rangle%
\end{array}%
\right) _{q},
\end{equation*}%
which gives%
\begin{eqnarray*}
\left( 
\begin{array}{c}
\left\vert 0\right\rangle \left\vert 0\right\rangle \\ 
\left\vert 0\right\rangle \left\vert 1\right\rangle \\ 
\left\vert 1\right\rangle \left\vert 0\right\rangle \\ 
\left\vert 1\right\rangle \left\vert 1\right\rangle%
\end{array}%
\right) _{p} &=&\left[ \frac{1}{\sqrt{2}}\left( 
\begin{array}{cccc}
1 & 0 & 0 & 1 \\ 
1 & 0 & 0 & -1 \\ 
0 & 1 & 1 & 0 \\ 
0 & -1 & 1 & 0%
\end{array}%
\right) \right] ^{-1}\left( 
\begin{array}{cccc}
1 & 0 & 0 & 1 \\ 
0 & 1 & 1 & 0 \\ 
1 & 0 & 0 & -1 \\ 
0 & 1 & -1 & 0%
\end{array}%
\right) \left( 
\begin{array}{c}
\left\vert 0\right\rangle \left\vert 0\right\rangle \\ 
\left\vert 0\right\rangle \left\vert 1\right\rangle \\ 
\left\vert 1\right\rangle \left\vert 0\right\rangle \\ 
\left\vert 1\right\rangle \left\vert 1\right\rangle%
\end{array}%
\right) _{q} \\
&=&\frac{1}{\sqrt{2}}\left( 
\begin{array}{cccc}
1 & 1 & 0 & 0 \\ 
0 & 0 & 1 & -1 \\ 
0 & 0 & 1 & 1 \\ 
1 & -1 & 0 & 0%
\end{array}%
\right) \frac{1}{\sqrt{2}}\left( 
\begin{array}{cccc}
1 & 0 & 0 & 1 \\ 
0 & 1 & 1 & 0 \\ 
1 & 0 & 0 & -1 \\ 
0 & 1 & -1 & 0%
\end{array}%
\right) \left( 
\begin{array}{c}
\left\vert 0\right\rangle \left\vert 0\right\rangle \\ 
\left\vert 0\right\rangle \left\vert 1\right\rangle \\ 
\left\vert 1\right\rangle \left\vert 0\right\rangle \\ 
\left\vert 1\right\rangle \left\vert 1\right\rangle%
\end{array}%
\right) _{q} \\
&=&\frac{1}{2}\left( 
\begin{array}{cccc}
1 & 1 & 0 & 0 \\ 
0 & 0 & 1 & -1 \\ 
0 & 0 & 1 & 1 \\ 
1 & -1 & 0 & 0%
\end{array}%
\right) \left( 
\begin{array}{cccc}
1 & 0 & 0 & 1 \\ 
0 & 1 & 1 & 0 \\ 
1 & 0 & 0 & -1 \\ 
0 & 1 & -1 & 0%
\end{array}%
\right) \left( 
\begin{array}{c}
\left\vert 0\right\rangle \left\vert 0\right\rangle \\ 
\left\vert 0\right\rangle \left\vert 1\right\rangle \\ 
\left\vert 1\right\rangle \left\vert 0\right\rangle \\ 
\left\vert 1\right\rangle \left\vert 1\right\rangle%
\end{array}%
\right) _{q}.
\end{eqnarray*}%
Therefore, 'momentum' basis is given by the transformation of the 'position'
basis as%
\begin{equation*}
\left( 
\begin{array}{c}
\left\vert 0\right\rangle \left\vert 0\right\rangle \\ 
\left\vert 0\right\rangle \left\vert 1\right\rangle \\ 
\left\vert 1\right\rangle \left\vert 0\right\rangle \\ 
\left\vert 1\right\rangle \left\vert 1\right\rangle%
\end{array}%
\right) _{p}=\frac{1}{2}\left( 
\begin{array}{cccc}
1 & 1 & 1 & 1 \\ 
1 & -1 & 1 & -1 \\ 
1 & 1 & -1 & -1 \\ 
1 & -1 & -1 & 1%
\end{array}%
\right) \left( 
\begin{array}{c}
\left\vert 0\right\rangle \left\vert 0\right\rangle \\ 
\left\vert 0\right\rangle \left\vert 1\right\rangle \\ 
\left\vert 1\right\rangle \left\vert 0\right\rangle \\ 
\left\vert 1\right\rangle \left\vert 1\right\rangle%
\end{array}%
\right) _{q}.
\end{equation*}

\subsection{Three-Qubit Entangled Basis}

As in the two-qubit, the product state in the 'momentum' basis is expanded
in terms of the 'position' basis%
\begin{eqnarray*}
&&\left\vert p^{\prime }\right\rangle \left\vert p^{\prime \prime
}\right\rangle \left\vert p^{\prime \prime \prime }\right\rangle \\
&=&\frac{1}{N^{\frac{3}{2}}}\sum\limits_{q^{\prime }}\exp \left( -\frac{i}{%
\hbar }p^{\prime }\cdot q^{\prime }\right) \ \left\vert q^{\prime
}\right\rangle \sum\limits_{q^{\prime \prime }}\exp \left( -\frac{i}{\hbar }%
p^{\prime \prime }\cdot q^{\prime \prime }\right) \ \left\vert q^{\prime
\prime }\right\rangle \sum\limits_{q^{\prime \prime \prime }}\exp \left( -%
\frac{i}{\hbar }p^{\prime \prime \prime }\cdot q^{\prime \prime \prime
}\right) \ \left\vert q^{\prime \prime \prime }\right\rangle \\
&=&\frac{1}{N^{\frac{3}{2}}}\sum\limits_{q^{\prime },q^{\prime \prime
},q^{\prime \prime \prime }}\exp \left( -\frac{i}{\hbar }p^{\prime }\cdot
q^{\prime }\right) \exp \left( -\frac{i}{\hbar }p^{\prime \prime }\cdot
q^{\prime \prime }\right) \exp \left( -\frac{i}{\hbar }p^{\prime \prime
\prime }\cdot q^{\prime \prime \prime }\right) \ \ \left\vert q^{\prime
}\right\rangle \ \left\vert q^{\prime \prime }\right\rangle \left\vert
q^{\prime \prime \prime }\right\rangle .
\end{eqnarray*}%
We can expressed the right hand side in terms of 'correlated' (entangled)
product states by writing, $q^{\prime \prime }=q^{\prime }+m$, $q^{\prime
\prime \prime }=q^{\prime }+l$ where $m$ and $l$ are the quantum-correlation
'distances', 
\begin{eqnarray*}
\left\vert p^{\prime }\right\rangle \left\vert p^{\prime \prime
}\right\rangle \left\vert p^{\prime \prime \prime }\right\rangle &=&\frac{1}{%
N^{\frac{3}{2}}}\sum\limits_{q^{\prime },m,l}\exp \left( -\frac{i}{\hbar }%
p^{\prime }\cdot q^{\prime }\right) \exp \left( -\frac{i}{\hbar }p^{\prime
\prime }\cdot \left( q^{\prime }+m\right) \right) \\
&&\times \exp \left( -\frac{i}{\hbar }p^{\prime \prime \prime }\cdot \left(
q^{\prime }+l\right) \right) \ \left\vert q^{\prime }\right\rangle \
\left\vert q^{\prime }+m\right\rangle \left\vert q^{\prime }+l\right\rangle
\\
&=&\frac{1}{N^{\frac{3}{2}}}\sum\limits_{q^{\prime },m,l}\exp \left( -\frac{i%
}{\hbar }\left( p^{\prime }+p^{\prime \prime }+p^{\prime \prime \prime
}\right) \cdot q^{\prime }\right) \\
&&\times \exp \left( -\frac{i}{\hbar }p^{\prime \prime }\cdot m\right) \exp
\left( -\frac{i}{\hbar }p^{\prime \prime \prime }\cdot l\right) \ \left\vert
q^{\prime }\right\rangle \ \left\vert q^{\prime }+m\right\rangle \left\vert
q^{\prime }+l\right\rangle .
\end{eqnarray*}%
Now writing $p^{\prime }+p^{\prime \prime }+p^{\prime \prime \prime }=p$, we
have%
\begin{eqnarray*}
\left\vert p^{\prime }\right\rangle \left\vert p-p^{\prime }-p^{\prime
\prime \prime }\right\rangle \left\vert p^{\prime \prime \prime
}\right\rangle &=&\frac{1}{\left( N\right) ^{\frac{3}{2}}}%
\sum\limits_{m,l}\exp \left( -\frac{i}{\hbar }\left( p-p^{\prime }\right)
\cdot m\right) \exp \left( -\frac{i}{\hbar }p^{\prime \prime \prime }\cdot
\left( l-m\right) \right) \\
&&\times \sum\limits_{q^{\prime }}\exp \left( -\frac{i}{\hbar }p\cdot
q^{\prime }\right) \left\vert q^{\prime }\right\rangle \ \left\vert
q^{\prime }+m\right\rangle \left\vert q^{\prime }+l\right\rangle \  \\
&=&\frac{1}{N}\sum\limits_{m,l}\exp \left( -\frac{i}{\hbar }\left(
p-p^{\prime }\right) \cdot m\right) \exp \left( -\frac{i}{\hbar }p^{\prime
\prime \prime }\cdot \left( l-m\right) \right) \\
&&\times \left\vert \psi _{p,m,l}\right\rangle ,
\end{eqnarray*}%
where 
\begin{equation*}
\left\vert \psi _{p,m,l}\right\rangle =\frac{1}{\sqrt{N}}\sum\limits_{q^{%
\prime }}\exp \left( -\frac{i}{\hbar }p\cdot q^{\prime }\right) \left\vert
q^{\prime }\right\rangle \ \left\vert q^{\prime }+m\right\rangle \left\vert
q^{\prime }+l\right\rangle ,
\end{equation*}%
is the three-qubit correlated (entangled) state. Substituting $N=2$, and $%
p=\pi \hbar k$, we have%
\begin{equation*}
\left\vert \psi _{k,m,l}\right\rangle =\frac{1}{\sqrt{2}}\sum\limits_{q^{%
\prime }=0}^{1}\exp \left( -i\pi k\cdot q^{\prime }\right) \left\vert
q^{\prime }\right\rangle \ \left\vert q^{\prime }+m\right\rangle \left\vert
q^{\prime }+l\right\rangle ,
\end{equation*}%
and obtain%
\begin{equation*}
\left\vert \psi _{0,0,0}\right\rangle =\frac{1}{\sqrt{2}}\left( \left\vert
0\right\rangle \left\vert 0\right\rangle \left\vert 0\right\rangle
+\left\vert 1\right\rangle \left\vert 1\right\rangle \left\vert
1\right\rangle \right) =\Theta _{3}^{+},
\end{equation*}%
which is the well-known Greenberger-Horne-Zeilinger state, $\left\vert
GHZ\right\rangle $, which is the maximally entangled state of three qubits
since $m=0$, and $l=0$ (a complete correlation). We give the expressions for
the rest of $\left\vert \psi _{k,m,l}\right\rangle $,%
\begin{equation*}
\left\vert \psi _{0,0,1}\right\rangle =\frac{1}{\sqrt{2}}\left( \left\vert
0\right\rangle \left\vert 0\right\rangle \left\vert 1\right\rangle
+\left\vert 1\right\rangle \left\vert 1\right\rangle \left\vert
0\right\rangle \right) =\Gamma ^{+},
\end{equation*}%
\begin{equation*}
\left\vert \psi _{0,1,0}\right\rangle =\frac{1}{\sqrt{2}}\left( \left\vert
0\right\rangle \left\vert 1\right\rangle \left\vert 0\right\rangle
+\left\vert 1\right\rangle \left\vert 0\right\rangle \left\vert
1\right\rangle \right) =\Omega ^{+},
\end{equation*}%
\begin{equation*}
\left\vert \psi _{0,1,1}\right\rangle =\frac{1}{\sqrt{2}}\left( \left\vert
0\right\rangle \left\vert 1\right\rangle \left\vert 1\right\rangle
+\left\vert 1\right\rangle \left\vert 0\right\rangle \left\vert
0\right\rangle \right) =\Xi ^{+},
\end{equation*}%
\begin{equation*}
\left\vert \psi _{1,0,0}\right\rangle =\frac{1}{\sqrt{2}}\left( \left\vert
0\right\rangle \left\vert 0\right\rangle \left\vert 0\right\rangle
-\left\vert 1\right\rangle \left\vert 1\right\rangle \left\vert
1\right\rangle \right) =\Theta _{3}^{-},
\end{equation*}%
\begin{equation*}
\left\vert \psi _{1,0,0}\right\rangle =\frac{1}{\sqrt{2}}\left( \left\vert
0\right\rangle \left\vert 0\right\rangle \left\vert 0\right\rangle
+\left\vert 1\right\rangle \left\vert 1\right\rangle \left\vert
1\right\rangle \right) =\Theta _{3}^{+},
\end{equation*}%
\begin{equation*}
\left\vert \psi _{1,0,1}\right\rangle =\frac{1}{\sqrt{2}}\left( \left\vert
0\right\rangle \left\vert 0\right\rangle \left\vert 1\right\rangle
-\left\vert 1\right\rangle \left\vert 1\right\rangle \left\vert
0\right\rangle \right) =\Gamma ^{-},
\end{equation*}%
\begin{equation*}
\left\vert \psi _{1,1,0}\right\rangle =\frac{1}{\sqrt{2}}\left( \left\vert
0\right\rangle \left\vert 1\right\rangle \left\vert 0\right\rangle
-\left\vert 1\right\rangle \left\vert 0\right\rangle \left\vert
1\right\rangle \right) =\Omega ^{-},
\end{equation*}%
\begin{equation*}
\left\vert \psi _{1,1,1}\right\rangle =\frac{1}{\sqrt{2}}\left( \left\vert
0\right\rangle \left\vert 1\right\rangle \left\vert 1\right\rangle
-\left\vert 1\right\rangle \left\vert 0\right\rangle \left\vert
0\right\rangle \right) =\Xi ^{-}.
\end{equation*}

Just like the Bell basis, the set of three-qubit correlated states, $\left\{
\left\vert \psi _{k,m,l}\right\rangle \right\} $, is a complete and
orthonormal basis set, since each element is connected to the complete
orthonormal product states by unitary transformations. In what follows, we
will to $\left\{ \left\vert \psi _{k,m,l}\right\rangle \right\} $ as the
three-particle correlated basis (TPCB).

\begin{remark}
There is another three-qubit entangled state, the so-called $W$ state. A $W$
state is a collective spin state with one `excitation' as compared to a
coherent spin state which is fully aligned. With all but one spin aligned in
the x-direction, the state can be written:%
\begin{equation*}
\left\vert W\right\rangle =\frac{1}{\sqrt{N}}\left( \left\vert \downarrow
\uparrow \uparrow ...\uparrow \right\rangle _{1}+\left\vert \uparrow
\downarrow \uparrow ...\uparrow \right\rangle _{2}+....\left\vert \uparrow
\uparrow \uparrow ...\downarrow \right\rangle _{N}\right)
\end{equation*}%
The $\left\vert W\right\rangle $ maybe viewed as a 'zero momentum'
superposition of 'position coordinates' indicated by the position of an
excitation in an array of fully-aligned spin%
\begin{eqnarray*}
\left\vert p\right\rangle &=&\frac{1}{\sqrt{N}}\sum\limits_{i=1}^{N}e^{\frac{%
i}{\hbar }p\cdot R_{i}}\left\vert R_{i}\right\rangle \\
\left\vert W\right\rangle &=&\frac{1}{\sqrt{N}}\sum\limits_{i=1}^{N}e^{\frac{%
i}{\hbar }0\cdot R_{i}}\left\vert R_{i}\right\rangle =\frac{1}{\sqrt{N}}%
\sum\limits_{i=1}^{N}\left\vert R_{i}\right\rangle
\end{eqnarray*}%
where the 'position' coordinate, $\left\vert R_{i}\right\rangle $,
correspond to the different position of one-excitation, $\downarrow $, in a
spin array $\left\vert spin\ \downarrow array\right\rangle _{i}$ as
indicated above (more appropriately in a spin ring). However, the 'position'
coordinates used is only a small subspace of the complete and orthonormal
product Hilbert space of an array of spin, moreover $\left\vert
W\right\rangle $ corresponds only to $\left\vert p\right\rangle =\left\vert
p=0\right\rangle $, i.e., zero-phase superposition. In general there are
also $N$ values of $p$ producing other $\left\vert W\right\rangle $-type
entangled states of array of spins with one excitation. Furthermore,
corresponding to the superposition of position eigenstates to produce a
Schroedinger wavefunction, one can also form an arbitrary superposition of
spin array $\left\vert R_{i}\right\rangle $ to form other correlated
(entangled) state, for exmple,%
\begin{equation*}
\left\vert \Psi \right\rangle =\sum\limits_{i=1}^{N}\psi \left( R_{i}\right)
\left\vert R_{i}\right\rangle
\end{equation*}%
where $\psi \left( R_{i}\right) $ may no longer be obtained from
Schroedinger equation, but from some optimization criteria, say \ for
optimal condition of a universal cloning machine, which for three-particle
entangled state, in the form (Phys. Rev. A 57, 2368 (1998)%
\begin{equation*}
\left\vert \Psi \right\rangle =\sqrt{\frac{2}{3}}\left\vert 100\right\rangle
-\sqrt{\frac{1}{6}}\left\vert 010\right\rangle -\sqrt{\frac{1}{6}}\left\vert
001\right\rangle
\end{equation*}
\end{remark}

It is worth pointing out that the method of generating entangled states
using the product of 'momentum' basis states, $\left\vert p\right\rangle $,
in the manner given above automatically yield complete orthonormal entangled
basis states.

\subsection{A qubit Teleportation Using Three-Particle Entanglement}

Here we will used as the quantum channel shared by Alice and Bob to be the
GHZ state%
\begin{equation*}
\left\vert GHZ\right\rangle =\Theta _{3}^{+}=\frac{1}{\sqrt{2}}\left(
\left\vert 0\right\rangle _{1}\left\vert 0\right\rangle _{2}\left\vert
0\right\rangle _{3}+\left\vert 1\right\rangle _{1}\left\vert 1\right\rangle
_{2}\left\vert 1\right\rangle _{3}\right) .
\end{equation*}%
As before Alice wants to send to Bob the unknown state, $\left\vert \psi
\right\rangle =\alpha \left\vert 0\right\rangle _{U}+\beta \left\vert
1\right\rangle _{U}$. To set up the teleportation, let us assume that
particles $1$ and $2$ are kept by Alice and particle $3$ is sent to Bob. So,
Alice has three particles ($U$, the one she wants to teleport, and particles 
$1$ and $2$, two of the entangled three qubits), and Bob has one particle $3$%
. In the total system, the state of these four particles is given by%
\begin{eqnarray*}
\left\vert \psi \right\rangle \otimes \left\vert \Phi _{3}^{+}\right\rangle
&=&\left( \alpha \left\vert 0\right\rangle _{U}+\beta \left\vert
1\right\rangle _{U}\right) \otimes \frac{1}{\sqrt{2}}\left( \left\vert
0\right\rangle _{1}\left\vert 0\right\rangle _{2}\left\vert 0\right\rangle
_{3}+\left\vert 1\right\rangle _{1}\left\vert 1\right\rangle _{2}\left\vert
1\right\rangle _{3}\right) \\
&=&\frac{\alpha }{\sqrt{2}}\left( \left[ \left\vert 0\right\rangle
_{U}\left\vert 0\right\rangle _{1}\left\vert 0\right\rangle _{2}\right]
\left\vert 0\right\rangle _{3}+\left[ \left\vert 0\right\rangle
_{U}\left\vert 1\right\rangle _{1}\left\vert 1\right\rangle _{2}\right]
\left\vert 1\right\rangle _{3}\right) \\
&&+\frac{\beta }{\sqrt{2}}\left( \left[ \left\vert 1\right\rangle
_{U}\left\vert 0\right\rangle _{1}\left\vert 0\right\rangle _{2}\right]
\left\vert 0\right\rangle _{3}+\left[ \left\vert 1\right\rangle
_{U}\left\vert 1\right\rangle _{1}\left\vert 1\right\rangle _{2}\right]
\left\vert 1\right\rangle _{3}\right) ,
\end{eqnarray*}%
where we have enclosed in square bracket the particles belonging to Alice.
We will now express Alice particle states in terms of the TPCB basis. We
have 
\begin{equation*}
\left\vert 0\right\rangle _{U}\left\vert 0\right\rangle _{1}\left\vert
0\right\rangle _{2}=\frac{1}{\sqrt{2}}\left( \Theta _{3}^{+}+\Theta
_{3}^{-}\right) _{U12},
\end{equation*}%
\begin{equation*}
\left\vert 1\right\rangle _{U}\left\vert 1\right\rangle _{1}\left\vert
1\right\rangle _{2}=\frac{1}{\sqrt{2}}\left( \Theta _{3}^{+}-\Theta
_{3}^{-}\right) _{U12},
\end{equation*}%
\begin{equation*}
\left\vert 0\right\rangle _{U}\left\vert 1\right\rangle _{1}\left\vert
1\right\rangle _{2}=\frac{1}{\sqrt{2}}\left( \Xi ^{+}+\Xi ^{-}\right) _{U12},
\end{equation*}%
\begin{equation*}
\left\vert 1\right\rangle _{U}\left\vert 0\right\rangle _{1}\left\vert
0\right\rangle _{2}=\frac{1}{\sqrt{2}}\left( \Xi ^{+}-\Xi ^{-}\right) _{U12}.
\end{equation*}%
Therefore we can write%
\begin{eqnarray*}
\left\vert \psi \right\rangle \otimes \left\vert \Phi _{3}^{+}\right\rangle
&=&\frac{\alpha }{\sqrt{2}}\left( \left[ \frac{1}{\sqrt{2}}\left( \Theta
_{3}^{+}+\Theta _{3}^{-}\right) _{U12}\right] \left\vert 0\right\rangle _{3}+%
\left[ \frac{1}{\sqrt{2}}\left( \Xi ^{+}+\Xi ^{-}\right) _{U12}\right]
\left\vert 1\right\rangle _{3}\right) \\
&&+\frac{\beta }{\sqrt{2}}\left( \left[ \frac{1}{\sqrt{2}}\left( \Xi
^{+}-\Xi ^{-}\right) _{U12}\right] \left\vert 0\right\rangle _{3}+\left[ 
\frac{1}{\sqrt{2}}\left( \Theta _{3}^{+}-\Theta _{3}^{-}\right) _{U12}\right]
\left\vert 1\right\rangle _{3}\right) ,
\end{eqnarray*}%
\begin{eqnarray*}
\left\vert \psi \right\rangle \otimes \left\vert \Phi _{3}^{+}\right\rangle
&=&\frac{1}{2}\left( \Theta _{3}^{+}\right) _{U12}\left( \alpha \left\vert
0\right\rangle _{3}+\beta \left\vert 1\right\rangle _{3}\right) +\frac{1}{2}%
\left( \Xi ^{+}\right) _{U12}\left( \alpha \left\vert 1\right\rangle
_{3}+\beta \left\vert 0\right\rangle _{3}\right) \\
&&+\frac{1}{2}\left( \Theta _{3}^{-}\right) _{U12}\left( \alpha \left\vert
0\right\rangle _{3}-\beta \left\vert 1\right\rangle _{3}\right) +\frac{1}{2}%
\left( \Xi ^{-}\right) _{U12}\left( \alpha \left\vert 1\right\rangle
_{3}-\beta \left\vert 0\right\rangle _{3}\right) \\
&=&\frac{1}{2}\left( \Theta _{3}^{+}\right) _{U12}I\left( 
\begin{array}{c}
\alpha \\ 
\delta%
\end{array}%
\right) +\frac{1}{2}\left( \Xi ^{+}\right) _{U12}\sigma _{x}\left( 
\begin{array}{c}
\alpha \\ 
\delta%
\end{array}%
\right) \\
&&+\frac{1}{2}\left( \Theta _{3}^{-}\right) _{U12}\sigma _{z}\left( 
\begin{array}{c}
\alpha \\ 
\delta%
\end{array}%
\right) +\frac{1}{2}\left( \Xi ^{-}\right) _{U12}\left( -i\sigma _{y}\right)
\left( 
\begin{array}{c}
\alpha \\ 
\delta%
\end{array}%
\right) .
\end{eqnarray*}%
Hence, it follows that regardless of the unknown state $\left\vert \psi
\right\rangle $, by using the maximally-entangled quantum channel, $%
\left\vert GHZ\right\rangle $, Alice can still perform only four
measurements with outcomes equally likely with probability equal to $\frac{1%
}{4}$. After Alice measurement, Bob's particle $3$ will have been projected
to one of the four pure states, after which Bob has to perform the necessary
transformation to regain the original state that Alice has, as indicated in
the following table,

\begin{center}
\begin{tabular}{|l|l|l|}
\hline
Alice measurement & Bob's particle state & Bob's transformation \\ \hline
$\left( \Theta _{3}^{+}\right) _{U12}$ & $\left( \alpha \left\vert
0\right\rangle _{3}+\beta \left\vert 1\right\rangle _{3}\right) $ & $I$ \\ 
\hline
$\left( \Theta _{3}^{-}\right) _{U12}$ & $\left( \alpha \left\vert
0\right\rangle _{3}-\beta \left\vert 1\right\rangle _{3}\right) $ & $\sigma
_{z}$ \\ \hline
$\left( \Xi ^{+}\right) _{U12}$ & $\left( \alpha \left\vert 1\right\rangle
_{3}+\beta \left\vert 0\right\rangle _{3}\right) $ & $\sigma _{x}$ \\ \hline
$\left( \Xi ^{-}\right) _{U12}$ & $\left( \alpha \left\vert 1\right\rangle
_{3}-\beta \left\vert 0\right\rangle _{3}\right) $ & $i\sigma _{y}$ \\ \hline
\end{tabular}
\end{center}

To transmit Alice's classical (measurement) result to Bob, Alice needs a
two-bit classical channel to transmit which one of the four equally likely
results . This is in contrast with other proposed more complex teleportation
scheme using the same maximally entangled $\left\vert GHZ\right\rangle $
quantum channel which employs Bob as an ancilla to transmit the unknown
state to Cliff, and use two classical bit between Alice and Bob and one
classical bit between Bob and Cliff. Here, whether we use the Bell basis or
TPCB basis to transmit an unknown quantum state, only two-bit classical
channel is simply needed directly between the sending and receiving parties,
and without the use of intermediary ancilla and only one local measurement
is involved.

\subsection{Teleportation Using Three-Particle Entanglement and an Ancilla}

Instead of keeping the particles $1$ and $2$ as done above, Alice only keep
particle $1$ and send particle $2$ to Bob and particle $3$ to Cliff. The
idea here is that Alice will locally only measure one of the orthonormal
two-particle entangled Bell states, $\left\vert \Phi ^{\pm }\right\rangle $, 
$\left\vert \Psi ^{\pm }\right\rangle $, instead of one of the complete
orthonormal three-particle correlated (entangled) states given above. The
price to pay in this scheme is that there is a need to have two bits of
classical communication channel between Alice and Bob and one more bit of
classical communication channel between Bob and Cliff. Moreover, two local
measurements has to be done, one local measurement by Alice on her Bell
states and one local measurement by Bob on his qubit, to implement the
teleportation of one qubit from Alice to Cliff. Thus, Bob becomes an ancilla
in this teleportation scheme.

Again, using the Alice wants to send to Bob the unknown state, $\left\vert
\psi \right\rangle =\alpha \left\vert 0\right\rangle _{U}+\beta \left\vert
1\right\rangle _{U}$, using the $\left\vert GHZ\right\rangle $ shared
quantum channel. To set up the teleportation, Alice keep particle $1$ and
send particle $2$ to Bob, and particle $3$ to Cliff. So, Alice has two
particles ($U$, the unknown qubit she wants to teleport and particle $1$,
one of the entangled three $\left\vert GHZ\right\rangle $ qubits). In the
total system, the state of these four particles is given by%
\begin{eqnarray*}
\left\vert \psi \right\rangle \otimes \left\vert GHZ\right\rangle &=&\left(
\alpha \left\vert 0\right\rangle _{U}+\beta \left\vert 1\right\rangle
_{U}\right) \otimes \frac{1}{\sqrt{2}}\left( \left\vert 0\right\rangle
_{1}\left\vert 0\right\rangle _{2}\left\vert 0\right\rangle _{3}+\left\vert
1\right\rangle _{1}\left\vert 1\right\rangle _{2}\left\vert 1\right\rangle
_{3}\right) \\
&=&\frac{\alpha }{\sqrt{2}}\left( \left[ \left\vert 0\right\rangle
_{U}\left\vert 0\right\rangle _{1}\right] \left\vert 0\right\rangle
_{2}\left\vert 0\right\rangle _{3}+\left[ \left\vert 0\right\rangle
_{U}\left\vert 1\right\rangle _{1}\right] \left\vert 1\right\rangle
_{2}\left\vert 1\right\rangle _{3}\right) \\
&&+\frac{\beta }{\sqrt{2}}\left( \left[ \left\vert 1\right\rangle
_{U}\left\vert 0\right\rangle _{1}\right] \left\vert 0\right\rangle
_{2}\left\vert 0\right\rangle _{3}+\left[ \left\vert 1\right\rangle
_{U}\left\vert 1\right\rangle _{1}\right] \left\vert 1\right\rangle
_{2}\left\vert 1\right\rangle _{3}\right) ,
\end{eqnarray*}%
where the particles enclosed in square brackets belongs to Alice. Upon
changing the Alice product basis states to Bell basis, we have%
\begin{eqnarray*}
&&\left\vert \psi \right\rangle \otimes \left\vert GHZ\right\rangle \\
&=&\left( \alpha \left\vert 0\right\rangle _{U}+\beta \left\vert
1\right\rangle _{U}\right) \otimes \frac{1}{\sqrt{2}}\left( \left\vert
0\right\rangle _{1}\left\vert 0\right\rangle _{2}\left\vert 0\right\rangle
_{3}+\left\vert 1\right\rangle _{1}\left\vert 1\right\rangle _{2}\left\vert
1\right\rangle _{3}\right) \\
&=&\frac{\alpha }{\sqrt{2}}\left( \left[ \frac{1}{\sqrt{2}}\left( \left\vert
\Phi ^{+}\right\rangle _{U1}+\left\vert \Phi ^{-}\right\rangle _{U1}\right) %
\right] \left\vert 0\right\rangle _{2}\left\vert 0\right\rangle _{3}+\frac{1%
}{\sqrt{2}}\left( \left\vert \Psi ^{+}\right\rangle _{U1}+\left\vert \Psi
^{-}\right\rangle _{U1}\right) \left\vert 1\right\rangle _{2}\left\vert
1\right\rangle _{3}\right) \\
&&+\frac{\beta }{\sqrt{2}}\left( \left[ \frac{1}{\sqrt{2}}\left( \left\vert
\Psi ^{+}\right\rangle _{U1}-\left\vert \Psi ^{-}\right\rangle _{U1}\right) %
\right] \left\vert 0\right\rangle _{2}\left\vert 0\right\rangle _{3}+\left[ 
\frac{1}{\sqrt{2}}\left( \left\vert \Phi ^{+}\right\rangle _{U1}-\left\vert
\Phi ^{-}\right\rangle _{U1}\right) \right] \left\vert 1\right\rangle
_{2}\left\vert 1\right\rangle _{3}\right) ,
\end{eqnarray*}%
\begin{eqnarray*}
\left\vert \psi \right\rangle \otimes \left\vert GHZ\right\rangle &=&\left(
\alpha \left\vert 0\right\rangle _{U}+\beta \left\vert 1\right\rangle
_{U}\right) \otimes \frac{1}{\sqrt{2}}\left( \left\vert 0\right\rangle
_{1}\left\vert 0\right\rangle _{2}\left\vert 0\right\rangle _{3}+\left\vert
1\right\rangle _{1}\left\vert 1\right\rangle _{2}\left\vert 1\right\rangle
_{3}\right) \\
&=&\frac{1}{2}\left( 
\begin{array}{c}
\left\vert \Phi ^{+}\right\rangle _{U1}\left( \alpha \left\vert
0\right\rangle _{2}\left\vert 0\right\rangle _{3}+\beta \left\vert
1\right\rangle _{2}\left\vert 1\right\rangle _{3}\right) \\ 
+\left\vert \Psi ^{+}\right\rangle _{U1}\left( \alpha \left\vert
1\right\rangle _{2}\left\vert 1\right\rangle _{3}+\beta \left\vert
0\right\rangle _{2}\left\vert 0\right\rangle _{3}\right)%
\end{array}%
\right) \\
&&+\frac{1}{2}\left( 
\begin{array}{c}
\left\vert \Phi ^{-}\right\rangle _{U1}\left( \alpha \left\vert
0\right\rangle _{2}\left\vert 0\right\rangle _{3}-\beta \left\vert
1\right\rangle _{2}\left\vert 1\right\rangle _{3}\right) \\ 
+\left\vert \Psi ^{-}\right\rangle _{U1}\left( \alpha \left\vert
1\right\rangle _{2}\left\vert 1\right\rangle _{3}-\beta \left\vert
0\right\rangle _{2}\left\vert 0\right\rangle _{3}\right)%
\end{array}%
\right) .
\end{eqnarray*}%
A local measurement of one of the four Bell states, which have equal
probability equal to $\frac{1}{4}$, will project the joint state of the
particles $2$ and $3$ possessed by Bob and Cliff, respectively, into one of
the entangled states shown above. Assume that Alice local measurement yields 
$\left\vert \Phi ^{+}\right\rangle _{U1}$, which is communicated to Bob
through a classical two-bit channel. Then the state of particles $2$ and $3$
is%
\begin{equation*}
\left\vert \psi _{23}\right\rangle =\alpha \left\vert 0\right\rangle
_{2}\left\vert 0\right\rangle _{3}+\beta \left\vert 1\right\rangle
_{2}\left\vert 1\right\rangle _{3}.
\end{equation*}%
In order to effect a transfer of the unknown qubit to Cliff, Bob must now
unentangle his qubit from that of Cliff. To do this Bob has to make a local
measurement on his qubit $2$.

Suppose Bob's measurement apparatus collapses his qubit state to two
possible outcomes, namely $k_{1}$ and $k_{2}$. Then Bob can decompose his
incoming state to the new basis states, namely, $\left\vert
k_{1}\right\rangle $ and $\left\vert k_{2}\right\rangle $, and write the
unitary/orthogonal transformation from the old basis to the new basis as%
\begin{eqnarray*}
\left\vert 0\right\rangle _{2} &=&\sin \theta \ \left\vert
k_{1}\right\rangle +\cos \theta \ \left\vert k_{2}\right\rangle , \\
\left\vert 1\right\rangle _{2} &=&\cos \theta \ \left\vert
k_{1}\right\rangle -\sin \theta \ \left\vert k_{2}\right\rangle .
\end{eqnarray*}%
Then $\left\vert \psi _{23}\right\rangle $ in terms of Bob's new basis is%
\begin{eqnarray*}
\left\vert \psi _{23}\right\rangle &=&\left( \sin \theta \ \left\vert
k_{1}\right\rangle +\cos \theta \ \left\vert k_{2}\right\rangle \right)
\alpha \left\vert 0\right\rangle _{3}+\left( \cos \theta \ \left\vert
k_{1}\right\rangle -\sin \theta \ \left\vert k_{2}\right\rangle \right)
\beta \left\vert 1\right\rangle _{3} \\
&=&\sin \theta \ \alpha \left\vert 0\right\rangle _{3}\left\vert
k_{1}\right\rangle +\cos \theta \ \alpha \left\vert 0\right\rangle
_{3}\left\vert k_{2}\right\rangle +\cos \theta \ \beta \left\vert
1\right\rangle _{3}\left\vert k_{1}\right\rangle -\sin \theta \beta
\left\vert 1\right\rangle _{3}\ \left\vert k_{2}\right\rangle \\
&=&\left( \sin \theta \ \alpha \left\vert 0\right\rangle _{3}+\cos \theta \
\beta \left\vert 1\right\rangle _{3}\right) \left\vert k_{1}\right\rangle
+\left( \cos \theta \ \alpha \left\vert 0\right\rangle _{3}-\sin \theta
\beta \left\vert 1\right\rangle _{3}\right) \ \left\vert k_{2}\right\rangle .
\end{eqnarray*}%
In general, the new basis and the old basis are mutually-unbiased basis%
\footnote{%
Example of mutually-unbiased bases are the position basis states and
momentum basis states.}, which means $\sin \theta =\cos \theta $ or $\theta =%
\frac{\pi }{4}$. Then $\left\vert \psi _{23}\right\rangle $ becomes%
\begin{equation*}
\left\vert \psi _{23}\right\rangle =\left( \ \alpha \left\vert
0\right\rangle _{3}+\ \beta \left\vert 1\right\rangle _{3}\right) \frac{1}{%
\sqrt{2}}\left\vert k_{1}\right\rangle +\left( \ \alpha \left\vert
0\right\rangle _{3}-\beta \left\vert 1\right\rangle _{3}\right) \ \frac{1}{%
\sqrt{2}}\left\vert k_{2}\right\rangle .
\end{equation*}%
Whatever Bob's measurement outcome is, Bob has to communicate to Cliff his
result through a one-bit classical channel. If the outcome of Bob's local
measurement is $k_{1}$ then Cliff has the unknown qubit from Alice and do
nothing. On the other hand if Bob's measurement yields $k_{2}$, then Cliff
has to perform a $\sigma _{z}$ transformation on his qubit to obtain the
unknown qubit from Alice.

\subsection{Two-Qubit Teleportation Using Three-Particle Entanglement}

We will now show that a full use of the capability of the GHZ quantum
channel is achieved when teleporting two qubit of information. As before,
the quantum channel shared by Alice and Bob to be the GHZ state%
\begin{equation*}
\left\vert GHZ\right\rangle =\Theta _{3}^{+}=\frac{1}{\sqrt{2}}\left(
\left\vert 0\right\rangle _{1}\left\vert 0\right\rangle _{2}\left\vert
0\right\rangle _{3}+\left\vert 1\right\rangle _{1}\left\vert 1\right\rangle
_{2}\left\vert 1\right\rangle _{3}\right) .
\end{equation*}%
But now Alice wants to send an unknown two-particle bits (triplet) to Bob,
namely,%
\begin{equation*}
\left\vert \psi \right\rangle =\alpha \left\vert 0\right\rangle
_{U1}\left\vert 0\right\rangle _{U2}+\delta \left\vert 1\right\rangle
_{U1}\left\vert 1\right\rangle _{U2},
\end{equation*}%
which we may write in vector form as%
\begin{equation*}
\left\vert \psi \right\rangle =\left( 
\begin{array}{c}
\alpha \left\vert 0\right\rangle _{U1}\left\vert 0\right\rangle _{U2} \\ 
\delta \left\vert 1\right\rangle _{U1}\left\vert 1\right\rangle _{U2}%
\end{array}%
\right) \Rightarrow \left( 
\begin{array}{c}
\alpha \\ 
\delta%
\end{array}%
\right) ,
\end{equation*}%
where $\left\vert \alpha \right\vert ^{2}+\left\vert \delta \right\vert
^{2}=1$. To set up the two-bit teleportation, particle $1$ is kept by Alice
and particle $2$ are$3$ are sent to Bob. In the total system, the state of
these three particles is given by%
\begin{eqnarray*}
\left\vert \psi \right\rangle \otimes \left\vert GHZ\right\rangle &=&\left(
\alpha \left\vert 0\right\rangle _{U1}\left\vert 0\right\rangle _{U2}+\delta
\left\vert 1\right\rangle _{U1}\left\vert 1\right\rangle _{U2}\right) \\
&&\otimes \frac{1}{\sqrt{2}}\left( \left\vert 0\right\rangle _{1}\left\vert
0\right\rangle _{2}\left\vert 0\right\rangle _{3}+\left\vert 1\right\rangle
_{1}\left\vert 1\right\rangle _{2}\left\vert 1\right\rangle _{3}\right) \\
&=&\frac{1}{\sqrt{2}}\left( 
\begin{array}{c}
\alpha \left( \left\vert 0\right\rangle _{U1}\left\vert 0\right\rangle
_{U2}\left\vert 0\right\rangle _{1}\left\vert 0\right\rangle _{2}\left\vert
0\right\rangle _{3}+\left\vert 0\right\rangle _{U1}\left\vert 0\right\rangle
_{U2}\left\vert 1\right\rangle _{1}\left\vert 1\right\rangle _{2}\left\vert
1\right\rangle _{3}\right) \\ 
+\delta \left( \left\vert 1\right\rangle _{U1}\left\vert 1\right\rangle
_{U2}\left\vert 0\right\rangle _{1}\left\vert 0\right\rangle _{2}\left\vert
0\right\rangle _{3}+\left\vert 1\right\rangle _{U1}\left\vert 1\right\rangle
_{U2}\left\vert 1\right\rangle _{1}\left\vert 1\right\rangle _{2}\left\vert
1\right\rangle _{3}\right)%
\end{array}%
\right) .
\end{eqnarray*}%
Now we change basis using the complete orthonormal set of three-particle
correlated (entangled) states%
\begin{equation*}
\left\vert \psi _{0,0,0}\right\rangle =\frac{1}{\sqrt{2}}\left( \left\vert
0\right\rangle \left\vert 0\right\rangle \left\vert 0\right\rangle
+\left\vert 1\right\rangle \left\vert 1\right\rangle \left\vert
1\right\rangle \right) =\Theta _{3}^{+},
\end{equation*}%
\begin{equation*}
\left\vert \psi _{0,0,1}\right\rangle =\frac{1}{\sqrt{2}}\left( \left\vert
0\right\rangle \left\vert 0\right\rangle \left\vert 1\right\rangle
+\left\vert 1\right\rangle \left\vert 1\right\rangle \left\vert
0\right\rangle \right) =\Gamma ^{+},
\end{equation*}%
\begin{equation*}
\left\vert \psi _{0,1,0}\right\rangle =\frac{1}{\sqrt{2}}\left( \left\vert
0\right\rangle \left\vert 1\right\rangle \left\vert 0\right\rangle
+\left\vert 1\right\rangle \left\vert 0\right\rangle \left\vert
1\right\rangle \right) =\Omega ^{+},
\end{equation*}%
\begin{equation*}
\left\vert \psi _{0,1,1}\right\rangle =\frac{1}{\sqrt{2}}\left( \left\vert
0\right\rangle \left\vert 1\right\rangle \left\vert 1\right\rangle
+\left\vert 1\right\rangle \left\vert 0\right\rangle \left\vert
0\right\rangle \right) =\Xi ^{+},
\end{equation*}%
\begin{equation*}
\left\vert \psi _{1,0,0}\right\rangle =\frac{1}{\sqrt{2}}\left( \left\vert
0\right\rangle \left\vert 0\right\rangle \left\vert 0\right\rangle
-\left\vert 1\right\rangle \left\vert 1\right\rangle \left\vert
1\right\rangle \right) =\Theta _{3}^{-},
\end{equation*}%
\begin{equation*}
\left\vert \psi _{1,0,1}\right\rangle =\frac{1}{\sqrt{2}}\left( \left\vert
0\right\rangle \left\vert 0\right\rangle \left\vert 1\right\rangle
-\left\vert 1\right\rangle \left\vert 1\right\rangle \left\vert
0\right\rangle \right) =\Gamma ^{-},
\end{equation*}%
\begin{equation*}
\left\vert \psi _{1,1,0}\right\rangle =\frac{1}{\sqrt{2}}\left( \left\vert
0\right\rangle \left\vert 1\right\rangle \left\vert 0\right\rangle
-\left\vert 1\right\rangle \left\vert 0\right\rangle \left\vert
1\right\rangle \right) =\Omega ^{-},
\end{equation*}%
\begin{equation*}
\left\vert \psi _{1,1,1}\right\rangle =\frac{1}{\sqrt{2}}\left( \left\vert
0\right\rangle \left\vert 1\right\rangle \left\vert 1\right\rangle
-\left\vert 1\right\rangle \left\vert 0\right\rangle \left\vert
0\right\rangle \right) =\Xi ^{-},
\end{equation*}%
to obtain%
\begin{equation*}
\left\vert 0\right\rangle _{U1}\left\vert 0\right\rangle _{U2}\left\vert
0\right\rangle _{1}=\frac{1}{\sqrt{2}}\left( \Theta _{3}^{+}+\Theta
_{3}^{-}\right) ,
\end{equation*}%
\begin{equation*}
\left\vert 0\right\rangle _{U1}\left\vert 0\right\rangle _{U2}\left\vert
1\right\rangle _{1}=\frac{1}{\sqrt{2}}\left( \Gamma ^{+}+\Gamma ^{-}\right) ,
\end{equation*}%
\begin{equation*}
\left\vert 0\right\rangle _{U1}\left\vert 1\right\rangle _{U2}\left\vert
0\right\rangle _{1}=\frac{1}{\sqrt{2}}\left( \Omega ^{+}+\Omega ^{-}\right) ,
\end{equation*}%
\begin{equation*}
\left\vert 0\right\rangle _{U1}\left\vert 1\right\rangle _{U2}\left\vert
1\right\rangle _{1}=\frac{1}{\sqrt{2}}\left( \Xi ^{+}+\Xi ^{-}\right) ,
\end{equation*}%
\begin{equation*}
\left\vert 1\right\rangle _{U1}\left\vert 0\right\rangle _{U2}\left\vert
0\right\rangle _{1}=\frac{1}{\sqrt{2}}\left( \Xi ^{+}-\Xi ^{-}\right) ,
\end{equation*}%
\begin{equation*}
\left\vert 1\right\rangle _{U1}\left\vert 0\right\rangle _{U2}\left\vert
1\right\rangle _{1}=\frac{1}{\sqrt{2}}\left( \Omega ^{+}-\Omega ^{-}\right) ,
\end{equation*}%
\begin{equation*}
\left\vert 1\right\rangle _{U1}\left\vert 1\right\rangle _{U2}\left\vert
0\right\rangle _{1}=\frac{1}{\sqrt{2}}\left( \Gamma ^{+}-\Gamma ^{-}\right) ,
\end{equation*}%
\begin{equation*}
\left\vert 1\right\rangle _{U1}\left\vert 1\right\rangle _{U2}\left\vert
1\right\rangle _{1}=\frac{1}{\sqrt{2}}\left( \Theta _{3}^{+}-\Theta
_{3}^{-}\right) .
\end{equation*}%
Upon changing the basis of Alice particles, we have%
\begin{equation*}
\left\vert \psi \right\rangle \otimes \left\vert GHZ\right\rangle =\frac{1}{%
\sqrt{2}}\left( 
\begin{array}{c}
\alpha \left( \frac{1}{\sqrt{2}}\left( \Theta _{3}^{+}+\Theta
_{3}^{-}\right) \left\vert 0\right\rangle _{2}\left\vert 0\right\rangle _{3}+%
\frac{1}{\sqrt{2}}\left( \Gamma ^{+}+\Gamma ^{-}\right) \left\vert
1\right\rangle _{2}\left\vert 1\right\rangle _{3}\right) \\ 
+\delta \left( \frac{1}{\sqrt{2}}\left( \Gamma ^{+}-\Gamma ^{-}\right)
\left\vert 0\right\rangle _{2}\left\vert 0\right\rangle _{3}+\frac{1}{\sqrt{2%
}}\left( \Theta _{3}^{+}-\Theta _{3}^{-}\right) \left\vert 1\right\rangle
_{2}\left\vert 1\right\rangle _{3}\right)%
\end{array}%
\right) ,
\end{equation*}%
\begin{eqnarray*}
\left\vert \psi \right\rangle \otimes \left\vert GHZ\right\rangle &=&\frac{1%
}{2}\left( 
\begin{array}{c}
\Theta _{3}^{+}\left( \alpha \left\vert 0\right\rangle _{2}\left\vert
0\right\rangle _{3}+\delta \left\vert 1\right\rangle _{2}\left\vert
1\right\rangle _{3}\right) \\ 
+\Theta _{3}^{-}\left( \alpha \left\vert 0\right\rangle _{2}\left\vert
0\right\rangle _{3}-\delta \left\vert 1\right\rangle _{2}\left\vert
1\right\rangle _{3}\right) \\ 
+\Gamma ^{+}\left( \delta \left\vert 0\right\rangle _{2}\left\vert
0\right\rangle _{3}+\alpha \left\vert 1\right\rangle _{2}\left\vert
1\right\rangle _{3}\right) \\ 
+\Gamma ^{-}\left( -\delta \left\vert 0\right\rangle _{2}\left\vert
0\right\rangle _{3}+\alpha \left\vert 1\right\rangle _{2}\left\vert
1\right\rangle _{3}\right)%
\end{array}%
\right) \\
&=&\frac{1}{2}\left( 
\begin{array}{c}
\Theta _{3}^{+}I\left( 
\begin{array}{c}
\alpha \\ 
\delta%
\end{array}%
\right) \\ 
+\Theta _{3}^{-}\sigma _{z}\left( 
\begin{array}{c}
\alpha \\ 
\delta%
\end{array}%
\right) \\ 
+\Gamma ^{+}\sigma _{x}\left( 
\begin{array}{c}
\alpha \\ 
\delta%
\end{array}%
\right) \\ 
+\Gamma ^{-}\left( -i\sigma _{y}\right) \left( 
\begin{array}{c}
\alpha \\ 
\delta%
\end{array}%
\right)%
\end{array}%
\right) .
\end{eqnarray*}%
Thus, depending on the result of Alice measurement using the equi-probable,
with probability$=\frac{1}{4}$, three-particle entanglement basis, $\Theta
_{3}^{+}$, $\Theta _{3}^{-}$, $\Gamma ^{+}$, and $\Gamma ^{-}$, which is
communicated to Bob via classical two-bit channel, Bob will then use the
inverse transformation to recover the original triplet sent by Alice.

Alice can also send an unknown singlet to Bob, namely,%
\begin{equation*}
\left\vert \psi \right\rangle =\alpha \left\vert 0\right\rangle
_{U1}\left\vert 1\right\rangle _{U2}+\delta \left\vert 1\right\rangle
_{U1}\left\vert 0\right\rangle _{U2}.
\end{equation*}%
However, the quantum channel shared by Alice and Bob must now be chosen to
be given by 
\begin{equation*}
\left\vert Q_{ch}\right\rangle =\frac{1}{\sqrt{2}}\left( \left\vert
0\right\rangle _{1}\left\vert 1\right\rangle _{2}\left\vert 0\right\rangle
_{3}+\left\vert 1\right\rangle _{1}\left\vert 0\right\rangle _{2}\left\vert
1\right\rangle _{3}\right)
\end{equation*}%
instead of $\left\vert GHZ\right\rangle $ used above. To set up the two-bit
teleportation, particle $1$ is kept by Alice and particle $2$ are$3$ are
sent to Bob. In the total system, the state of these three particles is
given by%
\begin{eqnarray*}
&&\left\vert \psi \right\rangle \otimes \left\vert Q_{ch}\right\rangle \\
&=&\left\{ \alpha \left\vert 0\right\rangle _{U1}\left\vert 1\right\rangle
_{U2}+\delta \left\vert 1\right\rangle _{U1}\left\vert 0\right\rangle
_{U2}\right\} \otimes \left\{ \frac{1}{\sqrt{2}}\left( \left\vert
0\right\rangle _{1}\left\vert 1\right\rangle _{2}\left\vert 0\right\rangle
_{3}+\left\vert 1\right\rangle _{1}\left\vert 0\right\rangle _{2}\left\vert
1\right\rangle _{3}\right) \right\} \\
&=&\frac{1}{\sqrt{2}}\left( 
\begin{array}{c}
\alpha \left( \left\vert 0\right\rangle _{U1}\left\vert 1\right\rangle
_{U2}\left\vert 0\right\rangle _{1}\left\vert 1\right\rangle _{2}\left\vert
0\right\rangle _{3}+\left\vert 0\right\rangle _{U1}\left\vert 1\right\rangle
_{U2}\left\vert 1\right\rangle _{1}\left\vert 0\right\rangle _{2}\left\vert
1\right\rangle _{3}\right) \\ 
+\delta \left( \left\vert 1\right\rangle _{U1}\left\vert 0\right\rangle
_{U2}\left\vert 0\right\rangle _{1}\left\vert 1\right\rangle _{2}\left\vert
0\right\rangle _{3}+\left\vert 1\right\rangle _{U1}\left\vert 0\right\rangle
_{U2}\left\vert 1\right\rangle _{1}\left\vert 0\right\rangle _{2}\left\vert
1\right\rangle _{3}\right)%
\end{array}%
\right) .
\end{eqnarray*}%
Upon changing the basis to three-particle entanglement basis, we have%
\begin{eqnarray*}
&&\left\vert \psi \right\rangle \otimes \left\vert Q_{ch}\right\rangle \\
&=&\frac{1}{\sqrt{2}}\left( 
\begin{array}{c}
\alpha \left( \frac{1}{\sqrt{2}}\left( \Omega ^{+}+\Omega ^{-}\right)
\left\vert 1\right\rangle _{2}\left\vert 0\right\rangle _{3}+\frac{1}{\sqrt{2%
}}\left( \Xi ^{+}+\Xi ^{-}\right) \left\vert 0\right\rangle _{2}\left\vert
1\right\rangle _{3}\right) \\ 
+\delta \left( \frac{1}{\sqrt{2}}\left( \Xi ^{+}-\Xi ^{-}\right) \left\vert
1\right\rangle _{2}\left\vert 0\right\rangle _{3}+\frac{1}{\sqrt{2}}\left(
\Omega ^{+}-\Omega ^{-}\right) \left\vert 0\right\rangle _{2}\left\vert
1\right\rangle _{3}\right)%
\end{array}%
\right) \\
&=&\frac{1}{2}\left( 
\begin{array}{c}
\Omega ^{+}\left( \delta \left\vert 0\right\rangle _{2}\left\vert
1\right\rangle _{3}+\alpha \left\vert 1\right\rangle _{2}\left\vert
0\right\rangle _{3}\right) \\ 
+\Xi ^{+}\left( \alpha \left\vert 0\right\rangle _{2}\left\vert
1\right\rangle _{3}+\delta \left\vert 1\right\rangle _{2}\left\vert
0\right\rangle _{3}\right) \\ 
+\Omega ^{-}\left( -\delta \left\vert 0\right\rangle _{2}\left\vert
1\right\rangle _{3}+\alpha \left\vert 1\right\rangle _{2}\left\vert
0\right\rangle _{3}\right) \\ 
+\Xi ^{-}\left( \alpha \left\vert 0\right\rangle _{2}\left\vert
1\right\rangle _{3}-\delta \left\vert 1\right\rangle _{2}\left\vert
0\right\rangle _{3}\right)%
\end{array}%
\right) \\
&=&\frac{1}{2}\left( 
\begin{array}{c}
\Omega ^{+}\sigma _{x}\left( 
\begin{array}{c}
\alpha \\ 
\delta%
\end{array}%
\right) \\ 
+\Xi ^{+}I\left( 
\begin{array}{c}
\alpha \\ 
\delta%
\end{array}%
\right) \\ 
+\Omega ^{-}\left( -i\sigma _{y}\right) \left( 
\begin{array}{c}
\alpha \\ 
\delta%
\end{array}%
\right) \\ 
+\Xi ^{-}\sigma _{z}\left( 
\begin{array}{c}
\alpha \\ 
\delta%
\end{array}%
\right)%
\end{array}%
\right) .
\end{eqnarray*}%
Thus, depending on the result of Alice measurement using the equi-probable,
with probability$=\frac{1}{4}$, three-particle entanglement basis, $\Omega
^{+}$, $\Omega ^{-}$, $\Xi ^{+}$, and $\Xi ^{-}$, which is communicated to
Bob via classical two-bit channel, Bob will then use the inverse
transformation to recover the original singlet sent by Alice.

\section{MECHANICAL MODELING OF QUANTUM ENTANGLEMENTS}

Here, we now give a simple and pedagogical implementation of quantum
entanglement using a chain of mechanical inverters, see-saws or swings. This
mechanical model has been previously used by the author in investigating the
fundamental physical limits of computational processes \cite{buot}.

\subsection{Mechanical model for entangled two qubits}

\begin{figure}[h]
\centering
\includegraphics[width=5.0289in]{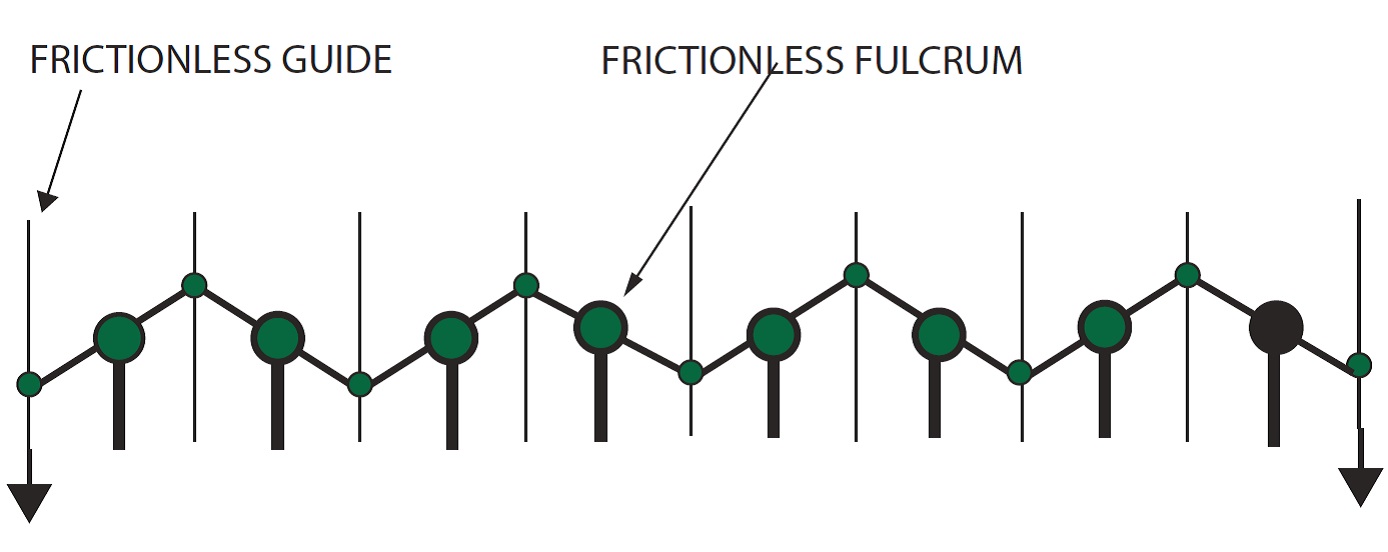}
\caption{An entangled spin triplet.}
\label{fig1}
\end{figure}

\begin{figure}[h]
\centering
\includegraphics[width=4.2748in]{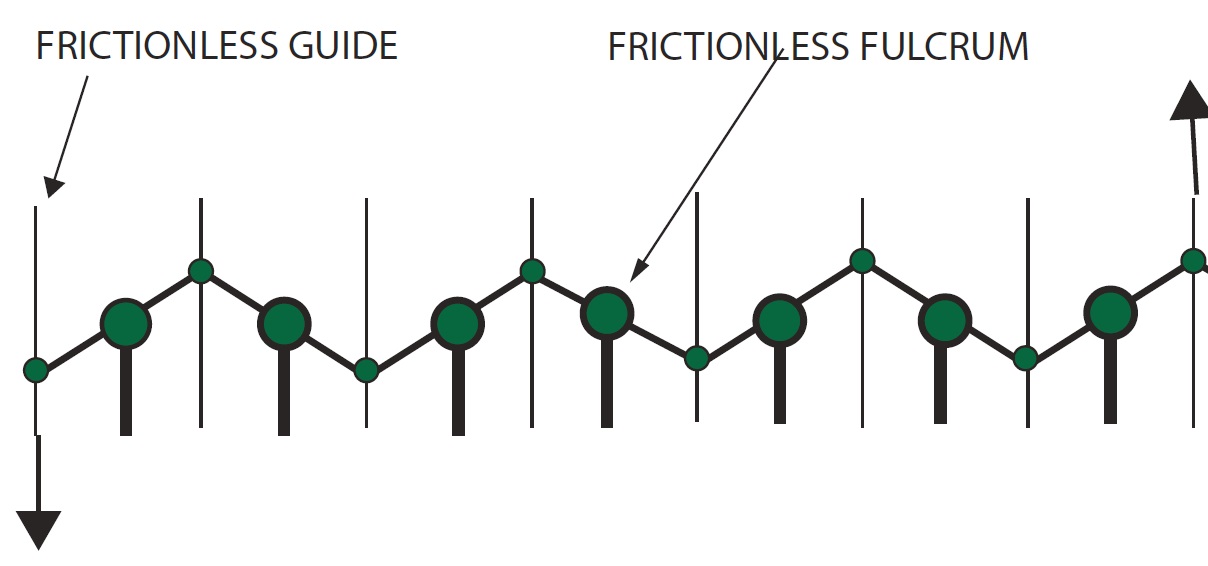}
\caption{An entangled spin singlet.}
\label{fig2}
\end{figure}

A schematic picture of an entangled two qubits is shown in Figs. \ref{fig1}
and \ref{fig2}. The frictionless guides and fulcrum would make the inverter
chain reversible. What is more important though, from the point of view of
entanglement, is the presence of rigid coupling. Rigid coupling allows for
simultaneous events to take place irrespective of the length of the
intermediary chain, thus capturing the essence of quantum entanglement.

No communication is involved since the whole process is a reconfiguration of
the whole chain system. No hidden variables and the deterministic
reconfiguration process does not violate Bell's inequality. In other words,
this mechanical model is conceptually an ideal model of quantum
entanglement. One is tempted to speculate whether spacetime provides rigid
coupling in realistic situations, indeed due to spacetime entanglement \cite%
{vix}.

Figure \ref{fig1} shows a configuration for a triplet entanglement.

Another independent configuration of the chain embodies the spin singlet
configuration shown in Fig. \ref{fig2}

\section{Diagrammatic Techniques}

In the following diagrammatic analyses, the inverter chains of Figs. \ref%
{fig1} and \ref{fig2} are abstracted into line segment representations.

\subsection{Two-level diagram representations of entangled two qubits}

\begin{figure}[h]
\centering
\includegraphics[width=5.9473in]{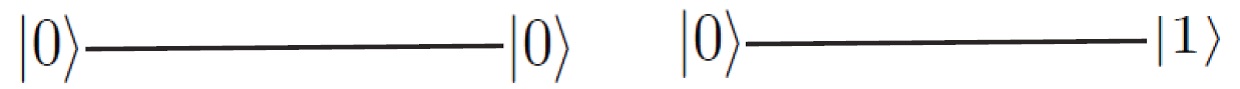}
\caption{Two two-state diagrams for an entangled two qubits. The
calculational entangled basis is obtained using Hadamard transformation or
discrete Fourier transform for two-state systems.}
\label{fig3}
\end{figure}

The line diagrams below is an abstractions of the inverter chain of Fig. \ref%
{fig1} and Fig. \ref{fig2}, respectively. The first line of Fig. \ref{fig3}
represents a two-state diagram, with entangled basis $\left\vert
0\right\rangle \left\vert 0\right\rangle $ and $\left\vert 1\right\rangle
\left\vert 1\right\rangle $. The second line represents a two-state diagram,
with entangled basis $\left\vert 0\right\rangle \left\vert 1\right\rangle $
and $\left\vert 1\right\rangle \left\vert 0\right\rangle $.

\subsubsection{Hadamard transformation for calculational entangled basis
states}

The Hadamard transform of the entangled basis states of two qubits given
above is what we referred to as the calculational entangled basis states.
These are the Bell entangled basis states derived formally before. For the
first diagram of Fig. \ref{fig3}, we have,%
\begin{equation*}
\left( 
\begin{array}{c}
\left\vert \Phi ^{+}\right\rangle \\ 
\left\vert \Phi ^{-}\right\rangle%
\end{array}%
\right) =\frac{1}{\sqrt{2}}\left( 
\begin{array}{cc}
1 & 1 \\ 
1 & -1%
\end{array}%
\right) \left( 
\begin{array}{c}
\left\vert 0\right\rangle \left\vert 0\right\rangle \\ 
\left\vert 1\right\rangle \left\vert 1\right\rangle%
\end{array}%
\right)
\end{equation*}%
and for the second diagram,%
\begin{equation*}
\left( 
\begin{array}{c}
\left\vert \Psi ^{+}\right\rangle \\ 
\left\vert \Psi ^{-}\right\rangle%
\end{array}%
\right) =\frac{1}{\sqrt{2}}\left( 
\begin{array}{cc}
1 & 1 \\ 
1 & -1%
\end{array}%
\right) \left( 
\begin{array}{c}
\left\vert 0\right\rangle \left\vert 1\right\rangle \\ 
\left\vert 1\right\rangle \left\vert 0\right\rangle%
\end{array}%
\right)
\end{equation*}

\subsection{Mechanical modeling of entangled three qubits}

\begin{figure}[h]
\centering
\includegraphics[width=2.6411in]{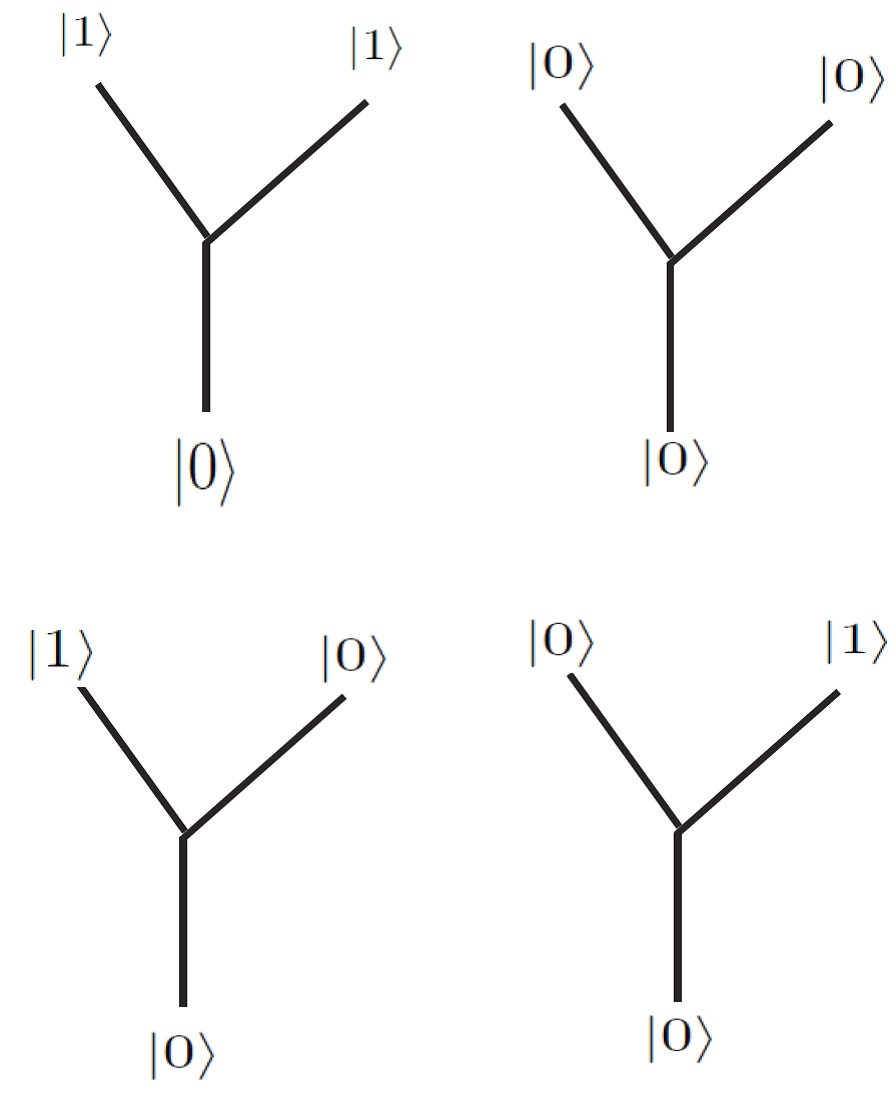}
\caption{Two-state four diagrams for entangled three qubits.}
\label{fig4}
\end{figure}

For entangled $n$ qubits, there are $\frac{2^{n}}{2}$ integer number of
two-state entangled basis diagrams. Thus, for three qubits the four diagrams
shown in Fig. \ref{fig4}.

From the diagrams of Fig. \ref{fig4}, and using the Hadamard transformation,
we can immediately write down the calculational entangled basis states for
three qubits.

\begin{equation*}
\left( 
\begin{array}{c}
\Xi ^{+} \\ 
\Xi ^{-}%
\end{array}%
\right) =\frac{1}{\sqrt{2}}\left( 
\begin{array}{cc}
1 & 1 \\ 
1 & -1%
\end{array}%
\right) \left( 
\begin{array}{c}
\left\vert 0\right\rangle \left\vert 1\right\rangle \left\vert 1\right\rangle
\\ 
\left\vert 1\right\rangle \left\vert 0\right\rangle \left\vert 0\right\rangle%
\end{array}%
\right)
\end{equation*}

\begin{equation*}
\left( 
\begin{array}{c}
\Theta _{3}^{+} \\ 
\Theta _{3}^{--}%
\end{array}%
\right) =\frac{1}{\sqrt{2}}\left( 
\begin{array}{cc}
1 & 1 \\ 
1 & -1%
\end{array}%
\right) \left( 
\begin{array}{c}
\left\vert 0\right\rangle \left\vert 0\right\rangle \left\vert 0\right\rangle
\\ 
\left\vert 1\right\rangle \left\vert 1\right\rangle \left\vert 1\right\rangle%
\end{array}%
\right)
\end{equation*}

\begin{equation*}
\left( 
\begin{array}{c}
\Omega ^{+} \\ 
\Omega ^{-}%
\end{array}%
\right) =\frac{1}{\sqrt{2}}\left( 
\begin{array}{cc}
1 & 1 \\ 
1 & -1%
\end{array}%
\right) \left( 
\begin{array}{c}
\left\vert 0\right\rangle \left\vert 1\right\rangle \left\vert 0\right\rangle
\\ 
\left\vert 1\right\rangle \left\vert 0\right\rangle \left\vert 1\right\rangle%
\end{array}%
\right)
\end{equation*}

\begin{equation*}
\left( 
\begin{array}{c}
\Gamma ^{+} \\ 
\Gamma ^{-}%
\end{array}%
\right) =\frac{1}{\sqrt{2}}\left( 
\begin{array}{cc}
1 & 1 \\ 
1 & -1%
\end{array}%
\right) \left( 
\begin{array}{c}
\left\vert 0\right\rangle \left\vert 0\right\rangle \left\vert 1\right\rangle
\\ 
\left\vert 1\right\rangle \left\vert 1\right\rangle \left\vert 0\right\rangle%
\end{array}%
\right)
\end{equation*}

\subsection{Mechanical modeling of entangled four qubits}

\begin{figure}[h]
\centering
\includegraphics[width=3.5362in]{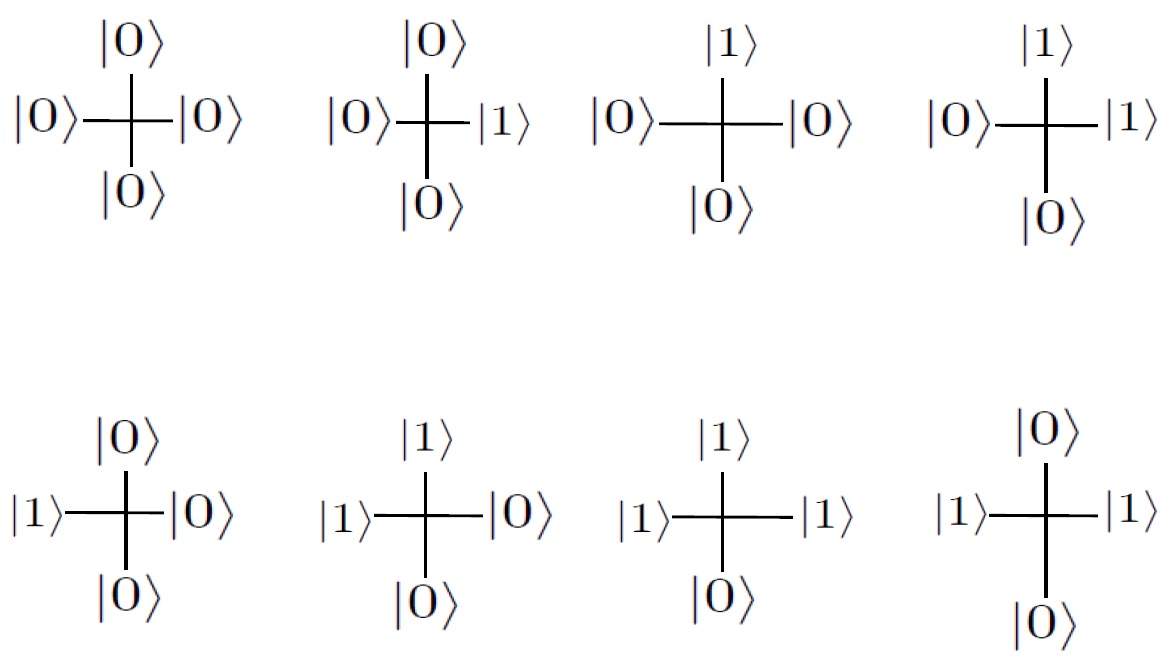}
\caption{Two-state eight diagrams for entangled four qubits.}
\label{fig5}
\end{figure}

We can consider eight two-entangled-state diagrams for entangled four
qubits. These are shown in Fig. \ref{fig5}

Following the same procedure using Hadamard Fourier transformation, we can
also immediately write down the calculational entangled basis states. Here,
we use our own notations, $\Phi _{j}^{\pm }$, for the transformed
calculational basis states. Following the order of diagrams in Fig. \ref%
{fig5}, we have%
\begin{equation*}
\left( 
\begin{array}{c}
\Phi _{1}^{+} \\ 
\Phi _{1}^{-}%
\end{array}%
\right) =\frac{1}{\sqrt{2}}\left( 
\begin{array}{cc}
1 & 1 \\ 
1 & -1%
\end{array}%
\right) \left( 
\begin{array}{c}
\left\vert 0\right\rangle \left\vert 0\right\rangle \left\vert
0\right\rangle \left\vert 0\right\rangle \\ 
\left\vert 1\right\rangle \left\vert 1\right\rangle \left\vert
1\right\rangle \left\vert 1\right\rangle%
\end{array}%
\right)
\end{equation*}%
\begin{equation*}
\left( 
\begin{array}{c}
\Phi _{2}^{+} \\ 
\Phi _{2}^{-}%
\end{array}%
\right) =\frac{1}{\sqrt{2}}\left( 
\begin{array}{cc}
1 & 1 \\ 
1 & -1%
\end{array}%
\right) \left( 
\begin{array}{c}
\left\vert 0\right\rangle \left\vert 0\right\rangle \left\vert
0\right\rangle \left\vert 1\right\rangle \\ 
\left\vert 1\right\rangle \left\vert 1\right\rangle \left\vert
1\right\rangle \left\vert 0\right\rangle%
\end{array}%
\right)
\end{equation*}%
\begin{equation*}
\left( 
\begin{array}{c}
\Phi _{3}^{+} \\ 
\Phi _{3}^{-}%
\end{array}%
\right) =\frac{1}{\sqrt{2}}\left( 
\begin{array}{cc}
1 & 1 \\ 
1 & -1%
\end{array}%
\right) \left( 
\begin{array}{c}
\left\vert 0\right\rangle \left\vert 0\right\rangle \left\vert
1\right\rangle \left\vert 0\right\rangle \\ 
\left\vert 1\right\rangle \left\vert 1\right\rangle \left\vert
0\right\rangle \left\vert 1\right\rangle%
\end{array}%
\right)
\end{equation*}%
\begin{equation*}
\left( 
\begin{array}{c}
\Phi _{4}^{+} \\ 
\Phi _{4}^{-}%
\end{array}%
\right) =\frac{1}{\sqrt{2}}\left( 
\begin{array}{cc}
1 & 1 \\ 
1 & -1%
\end{array}%
\right) \left( 
\begin{array}{c}
\left\vert 0\right\rangle \left\vert 0\right\rangle \left\vert
1\right\rangle \left\vert 1\right\rangle \\ 
\left\vert 1\right\rangle \left\vert 1\right\rangle \left\vert
0\right\rangle \left\vert 0\right\rangle%
\end{array}%
\right)
\end{equation*}%
\begin{equation*}
\left( 
\begin{array}{c}
\Phi _{5}^{+} \\ 
\Phi _{5}^{-}%
\end{array}%
\right) =\frac{1}{\sqrt{2}}\left( 
\begin{array}{cc}
1 & 1 \\ 
1 & -1%
\end{array}%
\right) \left( 
\begin{array}{c}
\left\vert 0\right\rangle \left\vert 1\right\rangle \left\vert
0\right\rangle \left\vert 0\right\rangle \\ 
\left\vert 1\right\rangle \left\vert 0\right\rangle \left\vert
1\right\rangle \left\vert 1\right\rangle%
\end{array}%
\right)
\end{equation*}%
\begin{equation*}
\left( 
\begin{array}{c}
\Phi _{6}^{+} \\ 
\Phi _{6}^{-}%
\end{array}%
\right) =\frac{1}{\sqrt{2}}\left( 
\begin{array}{cc}
1 & 1 \\ 
1 & -1%
\end{array}%
\right) \left( 
\begin{array}{c}
\left\vert 0\right\rangle \left\vert 1\right\rangle \left\vert
1\right\rangle \left\vert 0\right\rangle \\ 
\left\vert 1\right\rangle \left\vert 0\right\rangle \left\vert
0\right\rangle \left\vert 1\right\rangle%
\end{array}%
\right)
\end{equation*}%
\begin{equation*}
\left( 
\begin{array}{c}
\Phi _{7}^{+} \\ 
\Phi _{7}^{-}%
\end{array}%
\right) =\frac{1}{\sqrt{2}}\left( 
\begin{array}{cc}
1 & 1 \\ 
1 & -1%
\end{array}%
\right) \left( 
\begin{array}{c}
\left\vert 0\right\rangle \left\vert 1\right\rangle \left\vert
1\right\rangle \left\vert 1\right\rangle \\ 
\left\vert 1\right\rangle \left\vert 0\right\rangle \left\vert
0\right\rangle \left\vert 0\right\rangle%
\end{array}%
\right)
\end{equation*}%
\begin{equation*}
\left( 
\begin{array}{c}
\Phi _{8}^{+} \\ 
\Phi _{8}^{-}%
\end{array}%
\right) =\frac{1}{\sqrt{2}}\left( 
\begin{array}{cc}
1 & 1 \\ 
1 & -1%
\end{array}%
\right) \left( 
\begin{array}{c}
\left\vert 0\right\rangle \left\vert 1\right\rangle \left\vert
0\right\rangle \left\vert 1\right\rangle \\ 
\left\vert 1\right\rangle \left\vert 0\right\rangle \left\vert
1\right\rangle \left\vert 0\right\rangle%
\end{array}%
\right)
\end{equation*}%
In general, one can systematically write down the calculational entangled
basis states for any number, $n$, of qubits by using the physically
meaningful diagrammatic techniques prescribed here. One does not even need
to draw the diagrams themselves, but simply list the $\frac{2^{n}}{2}$
entangled basis basis to deduce the calculational entangled basis states by
Hadamard transformation of the two configurations of imagined entangled
diagrams.

\section{Concluding Remarks}

The apparently perfect scheme of the mechanical model in describing quantum
entanglement seems to advocate the existence of entangled spacetime-medium
to take the place of the chain of inverters in some unknown forms. An
interesting analogy is the bound/entangled pair of vortices and antivortex
on the surface of fluids which are really the ends of a vortex chain (tube)
forming a U-shape underneath the surface.  For example, in $3$-D the
superfluid quantized vortices form a metastable closed ring or open chain
ending at the surface. A vortex chain with both ends ending at the same
surface appears as a bound vortex/anti-vortex pair at the surface. Thus, it
seems extra dimensions are needed in spacetime to have a theory of
entanglement. The work of Ooguri \cite{tokyo} and collaborators shows that
this quantum entanglement generates the extra dimensions of the
gravitational theory. "It appears that it seems possible to generate a
geometric connection between entangled qubits, even though there is no
direct interaction between the two systems \cite{malcedona}. Furthermore,
the solid and reliable structure of spacetime is due to the ghostly features
of entanglement.

Could it be that besides the geometrical spacetime aspects of gravity, there
is a purely quantum mechanical aspect of spacetime geometry with extra
dimensons that give rise to entanglement? A hint along this idea is also
given by Malcedona \cite{malcedona} when he stated that "One can consider,
therefore, a pair of black holes where all the microstates are
\textquotedblleft entangled.\textquotedblright\ Namely, if we observe one of
the black holes in one particular microstate, then the other has to be in
exactly the same microstate. A pair of black holes in this particular EPR
entangled state would develop a wormhole, or Einstein-Rosen bridge,
connecting them through the inside. The geometry of this wormhole is given
by the fully extended Schwarzschild geometry. It is interesting that both
wormholes and entanglement naively appear to lead to a propagation of
signals faster than light." "It was known that quantum entanglement is
related to deep issues in the unification of general relativity and quantum
mechanics, such as the black hole information paradox and the firewall
paradox," says Hirosi Ooguri" \cite{tokyo}.

In any case, from the computational point of view, deriving the
computational entangled basis states have now been given for the first-time
a systematic and physically meaningful procedure using a diagrammatic
foundation. One can also define a closed group of entanglement diagrams. For
example, a direct product of entangled diagrams in Fig. 3 generates the
ordered fusion algebra: 
\begin{eqnarray*}
\text{triplet}\otimes \text{triplet} &=&\text{triplet} \\
\text{triplet}\otimes \text{singlet} &=&\text{singlet} \\
\text{singlet}\otimes \text{singlet} &=&\text{triplet}
\end{eqnarray*}

\end{document}